\documentclass[aps, onecolumn, superscriptaddress, 12pt]{revtex4}

\usepackage[english]{babel}
\usepackage{hyperref}
\usepackage{amsfonts}
\usepackage{multirow}
\usepackage{soul}

\usepackage[pdftex]{graphicx}
\usepackage{color}
\usepackage[toc,page]{appendix}
\usepackage{float}

\usepackage{amsmath, bbold}
\usepackage{amssymb}
\usepackage{mathtools}
\usepackage{braket}
\usepackage{cancel}

\usepackage{filecontents}
\usepackage[sort&compress]{natbib}

 \newcommand{\latdev}[1]{{\fontfamily{qcr}\selectfont #1}}

\begin{document}

\title{Testing quantum computers with the \\ protocol of quantum state matching}

\author{Adrian Ortega}
\email{ortega.adrian@wigner.hu}
\affiliation{Wigner  RCP,  Konkoly-Thege  M. u. 29-33, H-1121 Budapest, Hungary}

\author{Orsolya K\'alm\'an}
\email{kalman.orsolya@wigner.hu}
\affiliation{Wigner  RCP,  Konkoly-Thege  M. u. 29-33, H-1121 Budapest, Hungary}

\author{Tam\'as Kiss}
\email{kiss.tamas@wigner.hu}
\affiliation{Wigner  RCP,  Konkoly-Thege  M. u. 29-33, H-1121 Budapest, Hungary}

\date{\today}


\begin{abstract}
The presence of noise in quantum computers hinders their effective operation. Even though quantum error correction can theoretically remedy this problem, its practical realization is still a challenge. Testing and benchmarking noisy, intermediate-scale quantum (NISC) computers is therefore of high importance. Here, we suggest the application of the so-called quantum state matching protocol for testing purposes. This protocol was originally proposed to determine if an unknown quantum state falls in a prescribed neighborhood of a reference state. 
We decompose the unitary specific to the protocol and construct the quantum circuit implementing one step of the dynamics for different characteristic parameters of the scheme and present test results for two different IBM quantum computers. 
By comparing the experimentally obtained relative frequencies of success to the ideal success probability with a maximum statistical tolerance, we discriminate statistical errors from device specific ones.
For the characterization of noise, we also use the fact that while the output of the ideal protocol is insensitive to the internal phase of the input state, the actual implementation may lead to deviations. For systematically varied inputs we find that the device with the smaller quantum volume performs better on our tests than the one with larger quantum volume, while for random inputs they show a more similar performance. 

\end{abstract}

\maketitle

\section{Introduction}
\label{sec:Introduction}

The fields of Quantum Computation and Quantum Information have received a huge boost in the last years with the advent of ``public" quantum computation. Current devices can be accessed remotely, opening the possibility for the larger public to carry out experiments and to test them by running programs. Quantum computers (qcs) can be based on several different physical systems such as superconducting qubits~\cite{IBMqcs,rigettiqc,oxfordqc}, trapped ions~\cite{ionqqc}, photonic devices~\cite{quixqc} and neutral atoms~\cite{pasqalqc}. Given all these possibilities, questions, such as computational efficiency, error correction capability, stability and computational power start to become important matters for future applications. 

In order to discriminate among the different technologies or to decide the optimal domain of applicability of a given quantum computer
one needs to devise ``measure sticks" or benchmarks. In the current so-called Noisy Intermediate-Scale Quantum (NISQ) era~\cite{AspuruRMP2022}, the question of what a suitable benchmark is, becomes tricky because we are still dealing with ``unfinished" technologies: qcs that contain a lot of errors, do not support efficient error correction, have low computational power, among other missing traits that classical computers have already overcome~\cite{Hennessybook2017}. Indeed, an argument has been made about the current field of quantum computer benchmarking, stressing the point that we are still in the exploratory stage~\cite{BlumeOSTI2019}. The last few years have brought the arrival of the first quantum benchmarks, the most prominent ones being the Quantum Volume (QV)~\cite{MollQST2018,CrossPRA2019} and the $Q$-score~\cite{atosqscore}. Yet,
the field is only starting and there is still a long road ahead. 

Current quantum benchmarks can be divided roughly into two categories (not taking into account benchmarks 
related to temporal stability~\cite{DasguptaArxiv2020}): the first is based on randomized circuits such as randomized benchmarking~\cite{EmersonJOptB2005,KnillPRA2008,MagesanPRL2011,MagesanPRA2012}, the quantum volume ~\cite{CrossPRA2019}, or random circuits with a certain (mirror) structure~\cite{ProctorNatPhys2021}; the second is based on the successful achievement of certain hallmark protocols, such as the Bernstein-Vazirani algorithm and
Grover’s search~\cite{LinkePNAS2017, WrightNatCom2019, ZimborasQuantum2022}, the Bell test, the matrix inversion procedure or Schr{\"o}dinger's microscope~\cite{GilyenArxiv2021}, as well as algorithms used in quantum chemistry~\cite{McCaskeyNQI2019}. Benchmarks based on randomized circuits in which a probability distribution is sampled are in general a good starting point to test qcs, but they usually average out particular errors~\cite{GilyenArxiv2021}. These particular errors might become important, especially if the given quantum computer is used to perform a specific task.

Our work follows the approach of the second category, as it is based on the so-called quantum state matching protocol~\cite{KalmanPRA2018}. In a single step of this protocol a specific entangling operation is applied on a pair of qubits, both prepared in the same initial state. Then one member of the pair is measured and depending on the result of the measurement, the other member of the pair is post-selected or discarded. This procedure leads to a complex nonlinear transformation on the post-selected qubit. Note that a similar procedure is applied to realize the Schr\"{o}dinger microscope \cite{GilyenArxiv2021}. The transformation in our case is constructed in a way that it has two superattractive fixed points \cite{Milnorbook2011} (which correspond to orthogonal quantum states), with their respective basins of attraction being separated by a circle on the Bloch sphere. By iterating the protocol one can decide whether a given unknown quantum state falls in one of the basins of attraction, i.e., whether it is in the circle-shaped neighborhood of one of the superattractive states. The radius of the circle can be prescribed, in a given implementation of the scheme it determines the matrix elements of the entangling unitary, and also affects the probability of success of the protocol. 

In this paper, we show how one can employ this protocol to test qcs. In contrast to the nonlinear protocol realizing the Schr\"{o}dinger microscope ~\cite{GilyenArxiv2021}, where the dynamics does not possess any attractive cycles, thus all initial states are chaotic, in our scheme, the nonlinear transformation has superattractive fixed points and a tuneable success probability. In this way, the protocol itself can decrease initial noise, while such fluctuations may be enhanced in the case of the Schr\"{o}dinger microscope \cite{KalmanJRLR2018}. 

The paper is organised as follows. Section~\ref{sec:idealprot} describes the ideal protocol and introduces the specific entangling unitary that is involved in it. Then, in Sec.~\ref{sec:decunit} we determine the optimal decomposition of this unitary into a minimum number of programmable quantum gates. In Sec.~\ref{sec:fram} we describe the statistical framework used to test the quantum computers, while in Sec.~\ref{sec:implprot} we present and analyse the results obtained from real qcs. We compare results obtained by using systematically varied inputs (Sec.~\ref{sec:sysinput}) and randomly chosen inputs (Sec.~\ref{sec:randinput}), as well as results post-processed with readout error mitigation (Sec.~\ref{sec:errmit}). In Section~\ref{sec:Conclusions} we conclude and give an outlook on future directions. Appendix~\ref{sec:app2} presents the dates of the experiments discussed in the paper.
\section{The ideal protocol}
\label{sec:idealprot}
Errors in current NISQ computers lead to deviations from the desired pure output state of a quantum computation. These errors can be systematic, which do not necessarily change the purity of the qubit state, or random ones, leading to mixed outputs. One might wish to be able to decide whether such an output state is ``close enough" to a desired pure state.  The quantum state matching protocol was originally proposed for this task~\cite{KalmanPRA2018}: given a pure qubit state as a reference and a circular neighborhood around it, one can design a scheme which transforms the unknown state closer to the reference state if it was originally inside the prescribed circle, or otherwise to a state that is orthogonal to the reference state. Assuming that the unknown state is at hand in many copies, the scheme can be further iterated, and due to the superattractive nature of the transformation it can, after a few steps, match the unknown state to the reference or its orthogonal pair, which can then be discriminated.

The fast convergence of the above mentioned protocol is due to the nonlinear nature of the underlying quantum state transformation. The nonlinearity arises because one takes two copies of the unknown state $\ket{\Phi_{0}}$, then applies a specific entangling unitary (which is determined by the reference state and the radius of tolerance) and then measures one of the qubits. If the measurement result is $0$, then the other qubit undergoes a nonlinear transformation compared to its initial state. It has been proven in Ref.~\cite{KalmanPRA2018} that one can think of the entangling unitary as being composed of local rotations, which are determined by the position of the reference state on the Bloch sphere, and a two-qubit unitary, which is determined by the prescribed tolerance. A scheme containing solely this two-qubit unitary realizes quantum state matching to the reference state $\ket{0}$. In this work, we will focus on the implementation of this latter protocol, which we describe in what follows.  

Mathematically, a pure state of a two-level system can be written as

\begin{equation}
|\Phi_{0}\rangle=\mathcal{N}_{0} \left(|0\rangle + z|1\rangle\right),  
\label{eq:Phi0}
\end{equation}
where $z\in\mathbb{C}\cup\infty$ and $\mathcal{N}_0$ is a normalization factor. $|\Phi_{0}\rangle$ can be represented as a point on the Bloch (or the Riemann) sphere or, equivalently, it can be represented as a point on the complex plane, the two representations being related by the stereographic projection. Let us denote by $\epsilon$ the radius of the circular shaped tolerance region on the complex plane around the origin (representing the quantum state $\ket{0}$). 
Then, let us take two qubits, both prepared in the same intial state $|\Phi_0\rangle$, and apply the entangling unitary 

\begin{equation}
    U_\epsilon = \begin{pmatrix}
      \epsilon & -\frac{1}{\sqrt{2}}\sqrt{1-\epsilon^2} & \frac{1}{\sqrt{2}}\sqrt{1-\epsilon^2} & 0 \\
      0 & \frac{1}{\sqrt{2}} & \frac{1}{\sqrt{2}} & 0\\
      0 & 0 & 0 & 1\\
      \sqrt{1-\epsilon^2} & \frac{1}{\sqrt{2}}\epsilon & -\frac{1}{\sqrt{2}}\epsilon & 0
    \end{pmatrix},
    \quad
    \text{where}
    \quad
    0<\epsilon\leq 1.
    \label{eq:Ueps}
\end{equation}

It can be easily seen that if one measures the second qubit to be in state $|0\rangle$, then the state of the first qubit can be written as 

\begin{equation}
    |\Phi_1 \rangle = \mathcal{N}_{1}\left(|0\rangle + f(z)|1\rangle\right),\quad \text{where} \quad f(z) = \frac{z^2}{\epsilon}.
    \label{eq:fz}
\end{equation}

Note that this is a quadratic (nonlinear) transformation of the initial state $\ket{\Phi_{0}}$, represented by the complex number $z$. If one iterates this protocol, then initial states with $|z|<\epsilon$ will converge to $\ket{0}$, while those with $|z|>\epsilon$ will converge to $\ket{1}$, as  the $f(z)$ complex map has two (super)attractive fixed points: $0$ and $\infty$ (corresponing to the quantum states $\ket{0}$ and $\ket{1}$, respectively). The two regions of convergence (the so-called Fatou set) are separated by a circle (the so-called Julia set) of radius $\epsilon$ containing points which do not converge, but evolve chaotically \cite{Milnorbook2011}, \cite{Beardonbook1991}. We note that superattractivity of the quantum states $\ket{0}$ and $\ket{1}$ is advantageous for the protocol, as it ensures the fastest possible convergence, yet, the closer the initial unknown state is to the $\epsilon$-circle, the more iterations are needed for the initial state to be matched with the reference state (with a given accuracy). 

\begin{figure}[htb]
  \centering
  \includegraphics[width=0.95\textwidth]{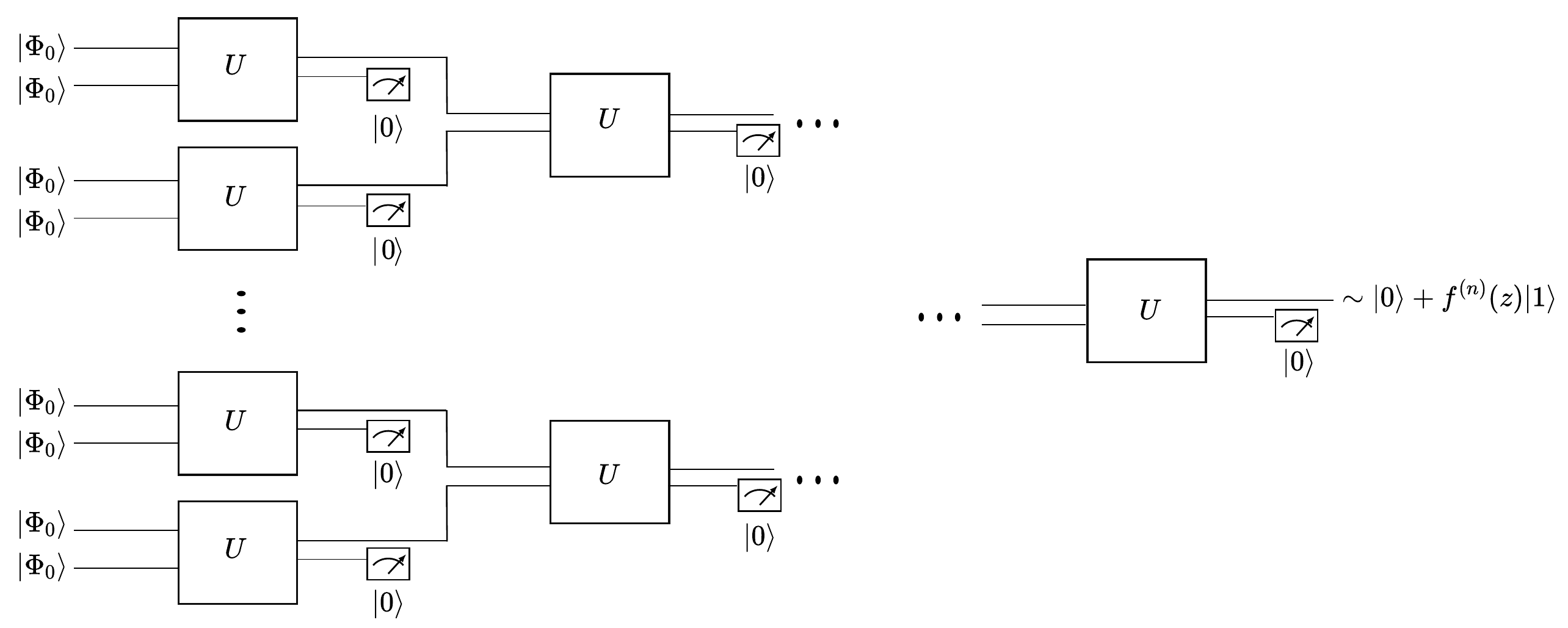}
  \caption{Quantum circuit for implementing $n$ iterations of the protocol. $f^{(n)}(z)$ denotes function composition
  of the complex map $f(z)$ of Eq.~(\ref{eq:fz}).}
  \label{fig:circ_full}
\end{figure}

In a quantum circuit the iteration of the protocol (and the corresponding complex map $f(z)$) can be carried out as depicted in Figure~\ref{fig:circ_full}. In order to carry out $n$ iterations, one needs $2^{n}$ qubits, all prepared in the same initial state $\ket{\Phi_{0}}$. For the first iteration, one forms $2^{n-1}$ pairs of these qubits and applies on every pair a $U_{\epsilon}$ gate. Every subsequent iterational step requires 2 times less $U_{\epsilon}$ gates compared to the previous step, so that in the complete circuit one needs to apply $\sum_{j=1}^{n} 2^{j-1}=2^{n}-1$ times the $U_{\epsilon}$ gate. Note that in the original proposal \cite{KalmanPRA2018}, one needs to perform measurements on one member of each pair and then post-select only the successfully transformed qubits in each step. As this so-called ``mid-circuit" measurement is not yet implemented in current commercially available quantum computers, one needs to postpone the post-selection step to the end of the quantum circuit, where all qubits are measured at once. 

Let us also note here that the quantum state matching protocol works in a practically analogous way for noisy inputs, i.e., when the copies of the initial state constitute a statistical mixture. In this case, the protocol can gradually decrease the noise and, with a good approximation, eventually purify the unknown state into the reference state or its orthogonal pair in a quite similar fashion to the case of pure inputs~\cite{KalmanPRA2018}.

In what follows, we look for an optimal decomposition of the unitary $U_{\epsilon}$ into elementary quantum gates, involving the lowest possible number of CNOT gates. We will show that in our case, two CNOT gates will suffice. Let us note that in the special case of $\epsilon=1$, which corresponds to the transformation $f(z)=z^2$, the protocol can also be realized with a single CNOT operation, however, the radius of tolerance is then the highest possible. In order to deviate from this rather trivial case, here, we will focus on cases with $\epsilon<1$.

\section{Decomposition of $U_\epsilon$ into programmable gates}
\label{sec:decunit}

In order to implement the protocol in a quantum computer, we first need to decompose $U_\epsilon$ (see Eq.~(\ref{eq:Ueps})) into elementary one- and two-qubit gates. It is known that any entangling two-qubit gate in $SU(4)$ can be decomposed into single-qubit gates and at most three CNOTs~\cite{KrausPRA2001,VidalPRA2004}. We will show that in our case, two CNOT gates will suffice. In order to find the decomposition, we closely follow the algorithmic-like procedure presented in~\cite{TucciArxiv2005}, which we briefly recall here. We note that similar decompositions can be found in Refs.~\cite{BullockPRA2003},\cite{VatanPRA2004}, where a further emphasis is made to determine the minimum number of gates needed from a given gate set. Here we do not restrict ourselves to such set of single-qubit gates as the actual physically implementable gates can differ in the different qcs. Instead, we accept that a further transpilation step will determine what native gates realize the single-qubit gates in our decomposition. 

As derived from Cartan's KAK decomposition by Khaneja et al.~\cite{KhanejaArxiv2000, KhanejaPRA2001}, any $U_{AB}\in SU(4)$ matrix can be decomposed as

\begin{equation}
  U_{AB} = (\mathcal{A}_1\otimes \mathcal{A}_2)e^{i\left(k_0+\vec{k}\cdot\vec{\Sigma}\right)}(\mathcal{B}_1\otimes \mathcal{B}_2).
  \label{eq:UAB}
\end{equation}

where $\mathcal{A}_l,\mathcal{B}_l\in SU(2)$ $(l=1,2)$ are local unitaries, $\vec{k}\in \mathbb{R}^{3}$, and 
\begin{equation}
    \vec{\Sigma}=\left(\sigma_{1}\otimes\sigma_{1},\sigma_{2}\otimes\sigma_{2}, \sigma_{3}\otimes\sigma_{3}\right)
\end{equation}
with $\sigma_{j}$ $(j=1,2,3)$ being the Pauli matrices. The entangling part of $U_{AB}$ is contained in the matrix

\begin{equation}
  U_{\rm ent} = \exp \left(i\vec{k}\cdot\vec{\Sigma} \right) =\exp\left(i\sum_{j=1}^3 k_j \sigma_j\otimes \sigma_j \right).
  \label{eq:UD}
\end{equation}
The procedure for the decomposition consists of the following steps.

\begin{enumerate}
\item Transform the unitary matrix $U_{AB}$ into the so-called magic basis using the matrix
\begin{equation}
  M = \frac{1}{\sqrt{2}}\begin{pmatrix}
    1 & 0 & 0 & i\\
    0 & i & 1 & 0\\
    0 & i & -1 & 0\\
    1 & 0 & 0 & -i\\
  \end{pmatrix}
\end{equation}
as
  \begin{equation}
    U' = M^\dagger U_{AB}M.
  \end{equation}
  
\item Separate $U'$ into its real ($U_{R}$) and imaginary parts ($U_{I}$). Here we note that $U_{R}$ and $U_{I}$ are real matrices (not unitary), which, due to the unitarity of $U'$, possess the properties $U_{I}U_{R}^{T}=U_{R}U_{I}^{T}$ and $U_{I}^{T}U_{R}=U_{R}^{T}U_{I}$. Consequently, according to a theorem by Eckart and Young \cite{EckartBAMS1939} one can find a pair of unitary matrices ($V_{A}$, $X_{A}$) with which a joint diagonalization of the $U_{R}$ and $U_{I}$ matrices is possible.

In order to do that one can first determine an SVD of $U_R$, namely
  \begin{equation}
    U_R = V_ADX_A^\dagger,
    \label{eq:UR}
  \end{equation}
where $D$ is a real diagonal matrix that contains the singular values (which are all non-negative) in its diagonal, while $V_A, X_A$ are unitary matrices. 

\item Convert the imaginary part $U_I$ using the SVD
  decomposition of $U_R$ as
  \begin{equation}    U'_I = V_A^\dagger U_I X_A.
  \label{eq:UI}
  \end{equation}
  Note that the proof of the theorem by Eckart and Young in Ref.~[\onlinecite{EckartBAMS1939}] also shows that $U'_{I}$ is Hermitian and commutes with $D$. 

\item Diagonalize $U'_I$ so that
\begin{align}
    U'_{I}=PGP^{\dagger}
    \label{eq:UIdiag}
\end{align}
with $G$ being the real diagonal
  matrix that contains the eigenvalues of $U'_{I}$, and $P$ being composed of the corresponding eigenvectors of $U'_{I}$. Since $D$ and $U'_I$ commute, $D$ also commutes with $P$ and $P^{\dagger}$. Thus
  we can write $U'$ as
  \begin{equation}
      U' =U_{R}+iU_{I} =V_{A}\left(D+iU'_{I}\right)X_{A}^{\dagger}= V_AP(D+iG)P^\dagger X_A^\dagger\\
      = Q_L(D+iG)Q_R^\dagger,
  \end{equation}
where we have defined $Q_L = V_AP$ and $Q_R=P^\dagger X_A^\dagger$. 

(Let us note here that in steps 2 and 3 the roles of the $U_{R}$ and $U_{I}$ matrices can be interchanged, i.e., one can first determine the SVD of $U_{I}$ and then transform $U_{R}$ and proceed along to diagonalize $U'$.)

\item Transform $U'$ back to the original basis. Note that $D+iG$ is a diagonal matrix and since $U'$ is unitary, $D+iG$ is also unitary, therefore the elements in the diagonal of $D+iG$ can be written as $e^{i\Phi_j}$ ($j=0,1,2,3$). The change to the original basis $M(D+iG)M^\dagger$ is equivalent to the following transformation of the $\Phi_{j}$ phases~\cite{TucciArxiv2005}

\begin{equation}
    (k_0,k_1,k_2,k_3)^T=\Lambda^{-1}(\Phi_0,\Phi_1,\Phi_2,\Phi_3)^T,
    \label{eq:kvsphi}
\end{equation}

where 

\begin{equation}
  \Lambda = \begin{pmatrix}
    1 & 1 & -1 & 1 \\
    1 & 1 & 1 & -1\\
    1 & -1 & -1 & -1\\
    1 & -1 & 1 & 1
  \end{pmatrix}.
\end{equation}
The local unitaries $\mathcal{A}_{l}$ and $\mathcal{B}_{l}$ of Eq.~(\ref{eq:UAB}) in the original basis are related to $Q_L$ and $Q_R$ via the identities
  \begin{equation}
    \begin{split}
      MQ_LM^\dagger &= \mathcal{A}_1\otimes \mathcal{A}_2,\\
      MQ_R^\dagger M^\dagger &= \mathcal{B}_1\otimes \mathcal{B}_2.\\
    \end{split}
    \label{eq:QRQL}
  \end{equation}
\end{enumerate}

In what follows we present the decomposition of the $U_{\epsilon}$ operation obtained by the above mentioned procedure. Since $U_\epsilon$ contains many zeros, one can find analytic expressions for all the terms in Eq.~\ref{eq:UAB}. 

Let us introduce the notation $\epsilon\equiv\cos\alpha$, ($\alpha\in[0,\pi/2)$) so that Eq.~(\ref{eq:Ueps}) becomes 

\begin{equation}
  U_\alpha =  \begin{pmatrix}
    \cos\alpha & -\frac{\sin\alpha}{\sqrt{2}} & \frac{\sin\alpha}{\sqrt{2}} & 0 \\
    0 & \frac{1}{\sqrt{2}} & \frac{1}{\sqrt{2}} & 0\\
    0 & 0 & 0 & 1\\
   \sin\alpha & \frac{\cos\alpha}{\sqrt{2}} & -\frac{\cos\alpha}{\sqrt{2}} & 0
  \end{pmatrix}.
\end{equation}
For a generic value of $\alpha$ (excluding the case $\alpha=\pi/4$, which, for simplicity, we do not detail here as it leads to a different decomposition), the singular values of the matrix $U_R$ (see Eq.~\ref{eq:UR}) are found to be
 
\begin{equation}
    \lambda_\pm = \frac{1}{8}\left(4-\sin(2\alpha) \pm r\right),
\end{equation}
where $r=\sqrt{8+\sin^{2}(2\alpha)}$.
We write $D$ as

\begin{equation}
  D = \operatorname{diag}(\lambda^{1/2}_+, \lambda^{1/2}_+, \lambda^{1/2}_-, \lambda^{1/2}_-),
  \label{eq:Dmat}
\end{equation}
and arrange in $X_{A}$ the corresponding $\vec{x}_j$ ($j=1,2,3,4$) right singular vectors as columns accordingly 

\begin{equation}
    X_A=
    \begin{pmatrix}
      y_1/N_1 &  0      & -1/N_1   &  0 \\
      0     & 1/N_2   &  0       & -y_2/N_2\\
      1/N_1 &  0      &  y_1/N_1 &  0\\
      0     & y_2/N_2 &  0       &  1/N_2\\
    \end{pmatrix}
    \label{eq:X_A}
\end{equation}

where  
\begin{align}
 y_1 &= \frac{3\sin(2\alpha) + r}{2\sqrt{2}\cos(2\alpha)}, \\
 y_2 &= \frac{\sin(2\alpha) - r}{2\sqrt{2}},
\end{align}

and $N_j = \sqrt{1+y_j^2}$ ($j=1,2$). (Note that one is free to choose the ordering of the singular values in $D$ as long as the corresponding right and left singular vectors are arranged accordingly in $X_{A}^{\dagger}$ and $V_{A}$.)

The matrix $V_A$ can be easily determined from Eq.~(\ref{eq:UR}) by computing $U_R\vec{x}_j = \lambda^{1/2}_{+/-}\vec{v}_j$ from which the left singular vectors $\vec{v}_j$ (the $j$th column of $V_A$) can be calculated for all $j$. Since $V_A$ is a unitary matrix, each of its columns are of the form $\vec{v}_j/M_j$, where $M_j=|\vec{v}_j|$.

Then, using the matrices $V_A$ and $X_A$ one finds from Eq.~(\ref{eq:UI}) that $U'_{I}$ has the form

\begin{equation}
  U'_{I}=  \frac{1}{2}\begin{pmatrix}
    0 & \tilde{r}/\tilde{M}_1 & 0 & 0\\
    \tilde{r}/\tilde{M}_1 & 0 & 0 & 0\\
    0 & 0 & 0 & r/\tilde{M}_2\\
    0 & 0 & r/\tilde{M}_2 &0
  \end{pmatrix},
  \label{eq:uip1}
\end{equation}
where
  \begin{align}
    \tilde{r} &= \frac{r}{2\sqrt{2}\cos(2\alpha)}\left[r^2-r\sin(2\alpha)-12\right],\notag\\
    \tilde{M}_1&=  2 \sqrt{2\lambda_+}N_1N_2M_2,\notag\\
    \tilde{M}_2&=  2\sqrt{\lambda_-}N_1N_2M_4.\notag
    \end{align}

The eigenvalues of $U'_I$ are easily obtained and thus $G$ can be given as (c.f. Eq.~(\ref{eq:UIdiag}))

\begin{equation}
  G = \frac{1}{2}\operatorname{diag}\left(\frac{\tilde{r}}{\tilde{M}_1}, -\frac{\tilde{r}}{\tilde{M}_1}, \frac{r}{\tilde{M}_2}, -\frac{r}
  {\tilde{M}_2}\right).
  \label{eq:Gmat}
\end{equation}

It is easy to see that the matrix $P$, which is composed of the eigenvectors of $U'_{I}$, can be given by $P = H\bigoplus H$, where $H$ is the $2\times 2$ Hadamard matrix.

Using Eqs.~(\ref{eq:Dmat}) and (\ref{eq:Gmat}) we can determine the phases $\Phi_{j}$ as

\begin{equation}
  \Phi_j = \operatorname{atan}\left(\frac{G_{jj}}{D_{jj}} \right),
\end{equation}
from which we can also calculate the phases $k_{j}$ using Eq.~(\ref{eq:kvsphi}). We find that $k_0 = k_1 = 0$, and only $k_2$ and $k_3$ are nonzero. 

Up to this point, we have determined the decomposition of $U_{\epsilon}$ into the entangling part $U_{\rm ent}$, and local parts $MQ_{L}M^{\dagger}$ and $MQ_{R}^{\dagger}M^{\dagger}$ as a function of $\alpha$ (or equivalently as a function of $\epsilon$). The local parts can further be decomposed into single-qubit transformations. Since  $Q_{L}=V_{A}P$ and $Q_{R}=P^{\dagger}X_{A}^{\dagger}$, we can look for the $\mathcal{A}_{1}$, $\mathcal{A}_{2}$, $\mathcal{B}_{1}$, $\mathcal{B}_{2}$ matrices as products of the decompositions of the respective constituting matrices, as e.g., $\mathcal{A}_{1}\otimes \mathcal{A}_{2}=MQ_{L}M^{\dagger}=MV_{A}PM^{\dagger}=\left(MV_{A}M^{\dagger}\right)\!\left(MPM^{\dagger}\right)$. Since $MPM^{\dagger}$ is block diagonal, it can easily be decomposed as
\begin{align}
    MPM^{\dagger}=-i\sigma_{2} \otimes S,
\end{align}
where
\begin{align}
    S=\frac{1}{\sqrt{2}}\begin{pmatrix}
      i & -1 \\
      1 & -i
    \end{pmatrix}.
\end{align}
In our case, $V_{A}$ and $X_{A}^{\dagger}$ are $SO(4)$ matrices of the form of Eq.~(\ref{eq:X_A}), i.e., they have nonzero elements in a chessboard pattern (there are entries only in the main diagonal and the $\pm2$-diagonals).
Moreover, $MV_{A}M^{\dagger}$ and $MX_{A}^{\dagger}M^{\dagger}$ are also $SO(4)$ matrices. It is well known that there exists a 2-1 homomorphism of $SU(2)\times SU(2) \rightarrow SO(4)$~\cite{TucciArxiv2005}, i.e., one can find two equivalent decompositions of an $SO(4)$ matrix. Due to the special form of the matrix $V_{A}$ ($X^{\dagger}_{A}$), we can find the decomposition by solving the system of equations represented by
\begin{align}
    MV_{A}M^{\dagger}=W_{1}\otimes W_{2}
\end{align}
where 
\begin{align}
    W_{i}=\begin{pmatrix}
      x_{i} & -y_{i}^{*} \\
      y_{i} & x_{i}^{*}
    \end{pmatrix},
    \quad (i=1,2), 
\end{align}
with $\left|x_{i}\right|^2+\left|y_{i}\right|^2=1$ (similarly for the case of $MX^{\dagger}_{A}M^{\dagger}$).

The local operations in the decomposition of $U_{\epsilon}$ can readily be implemented in quantum computers. $U_{\rm ent}$ however cannot be directly realized, thus one needs to look for an optimal decomposition for it in terms of programmable one- and two-qubit quantum gates. 

According to Theorem 2 of Ref.~[\onlinecite{VidalPRA2004}] $U_{\rm ent}$ can be realized with two CNOTs (plus some one-qubit gates) if it can be written as

\begin{equation}
    U = \exp\left[-i(h_1\sigma_1\otimes\sigma_1 + h_2\sigma_2\otimes\sigma_2)\right],\quad \text{where} \quad h_1\geq h_2 \geq 0.
    \label{eq:UDcan}
\end{equation}

In our case, for a generic value of $\epsilon$ we find that $k_1=0$, and $k_2,k_3<0$, so that Eq.~(\ref{eq:UD})  can be written as 

\begin{equation}
  U_{\rm ent}=\exp\left(i\sum_{j=1}^3 k_j \sigma_j\otimes \sigma_j\right) = \exp\left[-i(|k_{2}|\sigma_2\otimes \sigma_2 + |k_{3}|\sigma_3\otimes \sigma_3)\right],
  \label{eq:Uent}
\end{equation}
which is different from Eq.~(\ref{eq:UDcan}) in that (i) in the exponent the term with $\sigma_1\otimes \sigma_1$ is missing, while $\sigma_3\otimes \sigma_3$ is present, and (ii) a priori we do not know whether $|k_2|>|k_3|$ or $|k_3|>|k_2|$. Nonetheless, it is possible to transform $U_{\rm ent}$ to the form of $U$ by performing operations such as $[R^{\dagger}_{i}(\mu)]^{\otimes 2}U_{\rm ent}[R_{i}(\mu)]^{\otimes 2}$, where $\left[R_{i}(\mu)\right]^{\otimes 2}=R_i(\mu)\otimes R_i(\mu)$ denotes simultaneous local rotations of the form $R_i(\mu) = \exp(-i\mu \sigma_i)$. In this way one can change $\sigma_j\otimes\sigma_j$ to $\sigma_{k}\otimes\sigma_{k}$ in the desired way, which can be considered as a swapping of the corresponding  components of the vector $\vec{k}$. (We note here that this is analogous to saying that the canonical class vector $\vec{k}$ of $U_{\rm ent}$ is equivalent to the canonical class vector $\vec{h}$ of $U$ ~\cite{TucciArxiv2005}, resulting in the fact that the two unitaries have the same entangling power). The action of the aforementioned transformations can be seen by applying them on the terms of the Taylor expansion of the exponential in Eq.~(\ref{eq:Uent})
\begin{align}
    [R^{\dagger}_i&(\mu)]^{\otimes 2}\!\left(|k_{2}|\sigma_2\!\otimes \!\sigma_2 + |k_{3}|\sigma_3\!\otimes\!\sigma_3\right)^n\![R_i(\mu)]^{\otimes 2}\!=\!\! 
    \left[[R^{\dagger}_i(\mu)]^{\otimes 2}(|k_{2}|\sigma_2\!\otimes \!\sigma_2 +|k_{3}|\sigma_3\!\otimes\!\sigma_3) [R_i(\mu)]^{\otimes 2} \right]^{\!n} \notag\\ 
 &=\left[|k_{2}|\!\left(R^{\dagger}_i(\mu)\sigma_2 R_i(\mu)\right)\!\otimes\!\left(R_i^\dagger(\mu)\sigma_2 R_i(\mu)\right) + |k_{3}|\!\left(R_i^\dagger(\mu)\sigma_3 R_i(\mu)\right)\!\otimes\! \left(R_i^\dagger(\mu)\sigma_3 R_i(\mu)\right)\right]^n 
\end{align}
and then using the identities
\begin{align}
    e^{i\mu \sigma_{i}}\sigma_{j}e^{-i\mu\sigma_{i}}=\begin{cases}
    \quad \sigma_{j} & \text{if} \quad i=j \\
    \quad \cos(2\mu)\sigma_{j}-\sin(2\mu)\varepsilon_{ijk}\sigma_{k} & \text{if} \quad i\neq j
    \end{cases}
\end{align}
where $\varepsilon_{ijk}$ is the Levi-Civita symbol.

In order to see how $U_{\rm ent}$ can be brought to the form of Eq.~(\ref{eq:UDcan}) in our case, we need to treat the cases $|k_2|\geq |k_3|$ and $|k_3| > |k_2|$ separately. When $|k_2|\geq |k_3|$ we need to apply the simultaneous rotation $\left[R_3(3\pi/4)\right]^{\otimes 2}$
resulting in the change $(0,|k_2|,|k_3|) \rightarrow (|k_2|,0,|k_3|)$, followed by the application of $\left[R_1(3\pi/4)\right]^{\otimes 2}$ leading to $(|k_2|,0,|k_3|) \rightarrow (|k_2|, |k_3|,0)$, which is the desired order. In the case when $|k_3|>|k_2|$, we need to apply $\left[R_2(\pi/4)\right]^{\otimes 2}$, resulting in the change $(0,|k_2|,|k_3|) \rightarrow (|k_3|, |k_2|,0)$, which transforms $U_{\rm ent}$ to the required form. 

After bringing $U_{\rm ent}$ to the form of $U$, one can apply the results of Ref.~[\onlinecite{VidalPRA2004}] and decompose the transformed unitary into two CNOTs and local operations. Here we present the decomposition corresponding to the case when $|k_{2}|\geq |k_{3}|$ (the $|k_{3}|> |k_{2}|$ case follows similarly) 

\begin{align}
    [R^{\dagger}_1(3\pi/4)]^{\otimes 2}[R^{\dagger}_3(3\pi/4)]^{\otimes 2} U_{\rm ent}
&[R_3(3\pi/4)]^{\otimes 2}[R_1(3\pi/4)]^{\otimes 2}= \notag \\ 
    &(w\otimes w^\dagger)U_{\rm CNOT}(u_2\otimes v_2)U_{\rm CNOT}(w^\dagger\otimes w),
\end{align}

where $w = (\mathbb{1} -i\sigma_1)/2$, $u_2= \exp(-i|k_2|\sigma_1)$ and $v_2 = \exp(i|k_3|\sigma_3)$. 

\begin{figure}[htb]
  \centering
  \includegraphics[scale=1.6]{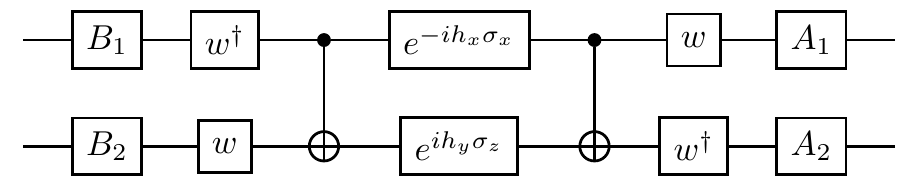}
  \caption{Circuit representation of $U_\epsilon$.}
  \label{fig:circ}
\end{figure}

The extra local rotations that we use to transform $U_{\rm ent}$ to the form of $U$ must be compensated in the full decomposition of $U_{\epsilon}$. This can be achieved by adding the respective inverse rotations, which will then multiply the local unitaries $\mathcal{A}_j$ and $\mathcal{B}_j$. We can thus define new local unitaries $A_j$ and $B_j$ as
\begin{equation}
\begin{split}
     A_1 \otimes A_2 &= (\mathcal{A}_1R_zR_x)\otimes (\mathcal{A}_2 R_zR_x),\\
     B_1 \otimes B_2 &= (R_xR_z\mathcal{B}_1)\otimes (R_x R_z\mathcal{B}_2),\\
\end{split}
\end{equation}
and write the final decomposition of $U_{\epsilon}$ containing only programmable quantum gates as 
(see also Fig.~\ref{fig:circ})
 
\begin{equation}
    U_\epsilon = (A_1\otimes A_2)(w\otimes w^\dagger)\operatorname{CNOT}(u_2\otimes v_2)\operatorname{CNOT}(w^\dagger\otimes w)(B_1 \otimes B_2).
 \end{equation}

In fact, in the actual implementation we aim to reduce the number of local unitaries, therefore we apply only a single unitary on every qubit before and after the central CNOTs (e.g., only $\bar{A}_{1}=A_{1}w$ acts on qubit 1  after the second CNOT). We note that the state preparation of the qubits is carried out before the application of $U_\epsilon$.

\section{Statistical framework for post-selection}
\label{sec:fram}

We are interested in a quantitative approach for testing real quantum computers. In the NISQ era, the impact of noise and errors in the devices is of special importance. Most of the time, it is not even clear how one can construct a theoretical model that takes into account all of them in a realistic way~\cite{SalonikArxiv2019}. Moreover, different quantum computers possess different sources of errors. For example, a great part of the errors in superconducting qcs comes from the readout step~\cite{MakhlinRMP2001,ChenPRA2019,AlexandrouArxiv2021,ZimborasQuantum2022}. On the other hand, in quantum computers based on cold atoms, a potential source of noise is the all-to-all qubit interaction~\cite{WrightNatCom2019}. Indeed, in~\cite{SalonikArxiv2019}  a wide spectrum of errors has been identified, ranging from environmental interactions, qubit interactions, imperfect operations and detection errors. All of them may be present in a quantum experiment in a nontrivial way. They can also vary in a complex manner as a function of time, even in the same device. If one would like to assess the performance of different quantum devices, a quantitative benchmark comes handy to compare the performance of various qcs. 

In our approach we will consider two quantities: the success probability, and the transformed quantum state after a given number of iterations. We will analyze the performance of the tested quantum computers by using the experimentally obtained relative frequencies to obtain corresponding quantities, which we compare to the ideal theoretical ones. 

Here we parameterize the input state with spherical Bloch-sphere coordinates $(\theta_{0},\phi_{0})$, which are related to the parameterization used in Eq.~(\ref{eq:Phi0}) by $z = e^{i\phi_{0}}\tan(\theta_{0}/2)$. Then the transformed state $\ket{\Phi_{n}}$ after $n$ steps of the protocol reads as
\begin{equation}
    \ket{\Phi_{n}}=\mathcal{N}_{n}\left[\epsilon^{2^{n}-1}\left(\cos\frac{\theta_0}{2}\right)^{2^{n}}\ket{0} + e^{i2^{n}\phi_{0}}\left(\sin\frac{\theta_0}{2}\right)^{2^{n}}\ket{1}\right].
    \label{eq:Phi_n}
\end{equation}
One can see that initial states with the same initial $\theta_{0}$ but different $\phi_{0}$ angle (i.e., states from a circle at a given latitude) will be mapped to states with the same $\theta_{n}$, i.e., to another circle at a different latitude. Consequently, if we measure the final kept qubit only in the computational basis, then, in an ideal case, only the value of $\theta_{0}$ should affect the measurement results. We will utilize this property of the protocol to test the performance of qcs.

The ideal (theoretical) success probability of performing $n$ iterations of the protocol can be expressed in the following way \cite{KalmanPRA2018}
\begin{equation}
    p_s^{(n)} 
    = \epsilon^{2^{n+1}-2}\left(\cos\frac{\theta_0}{2}\right)^{2^{n+1}} + \left(\sin\frac{\theta_0}{2}\right)^{2^{n+1}}
    =\frac{1}{\mathcal{N}_{n}^{2}}.
    \label{eq:ps}
\end{equation}

It is important to note here that $p_{s}^{(n)}$ is independent of the initial angle $\phi_{0}$, thus, in the case of an ideal (noise-free) quantum circuit, $p_{s}^{(n)}$ should not vary if we change $\phi_{0}$. We will use this fact as one of our test tools. 

Another aspect we aim at is to be able to distinguish statistical errors from device specific errors~\cite{GilyenArxiv2021}. The first type of errors are independent of the physical system, they are due to the finite number of experiments that can be realized in any quantum system~\cite{ParisLNP2004}. The second type of errors are those which are specific to the physical system itself (e.g., quantum gate errors and readout errors). From a benchmarking perspective, these latter are the ones we look for. Indeed, in~\cite{GilyenArxiv2021} it was pointed out that a quantitative benchmark score should behave as $\Delta_S/\sqrt{M} + \Delta_D$, where $M$ is the total number of experiments and $\Delta_S$, $\Delta_D$ are the statistical noise and the device noise, respectively. Thus, if $M$ is large enough, one should be able to detect and quantify device specific errors.

Since, for a given $\epsilon$, $p_s$ decreases double-exponentially with each iteration, in what follows, we will focus only on one step of the protocol. 
Let us denote the relative frequency of success after the first step with $p_s^{(e)}=N/M$, where $N$ is the number of favorable outcomes. In order to quantify statistical errors, we can introduce the usual sigma notation with the help of the standard deviation $\sigma = \sqrt{p_s(1-p_s)/M}$. We will mostly use $3 \sigma$ tolerance, meaning that more than $99\%$ of the experiments should fall in the interval $[p_{s}-3 \sigma , p_{s}+ 3 \sigma]$. This implies that relative frequencies outside this interval are caused by device errors with high probability.
In the next section, we investigate some quantitative properties of two IBM quantum computers using this method.  

\section{Implementation of the quantum state matching protocol}
\label{sec:implprot}

For the implementation of one step of the protocol we constructed the 2-qubit quantum circuit in Fig.~\ref{fig:circ} in two freely available IBM devices:  \latdev{ibmq\_manila} and \latdev{ibmq\_lima}. The reason for choosing these devices is their significantly different Quantum Volumes (QV) \cite{IBMqcs}: \latdev{ibmq\_manila} has a QV of 32, while \latdev{ibmq\_lima} has a QV of 8. The QV is a currently widely used measure of performance of quantum devices. Our purpose is to see if the differences suggested by the value of the QV are also reflected in the results of our tests with the quantum state matching protocol.

\subsection{Systematically varied inputs}
\label{sec:sysinput}

First, we investigate how the relative frequency of success $p_{s}^{(e)}$ in the different experiments compare to the ideal success probability. In our two-qubit circuit, $p_{s}^{(e)}$ can be determined by taking the counts corresponding to measuring the the states $\ket{00}$ and $\ket{10}$ at the end of the circuit. We note that in our implementations we always set the first qubit as the one which we want to transform, and the second as the one according to which we post-select, thus, cases, where the second qubit is measured to be $\ket{1}$, are not considered to be successful events. Furthermore, we also payed attention to the circuit topology: we always used neighboring qubits
and allowed for transpilation optimizations whenever possible. 

\begin{figure}[htb]
  \centering
  \includegraphics[scale=0.55]{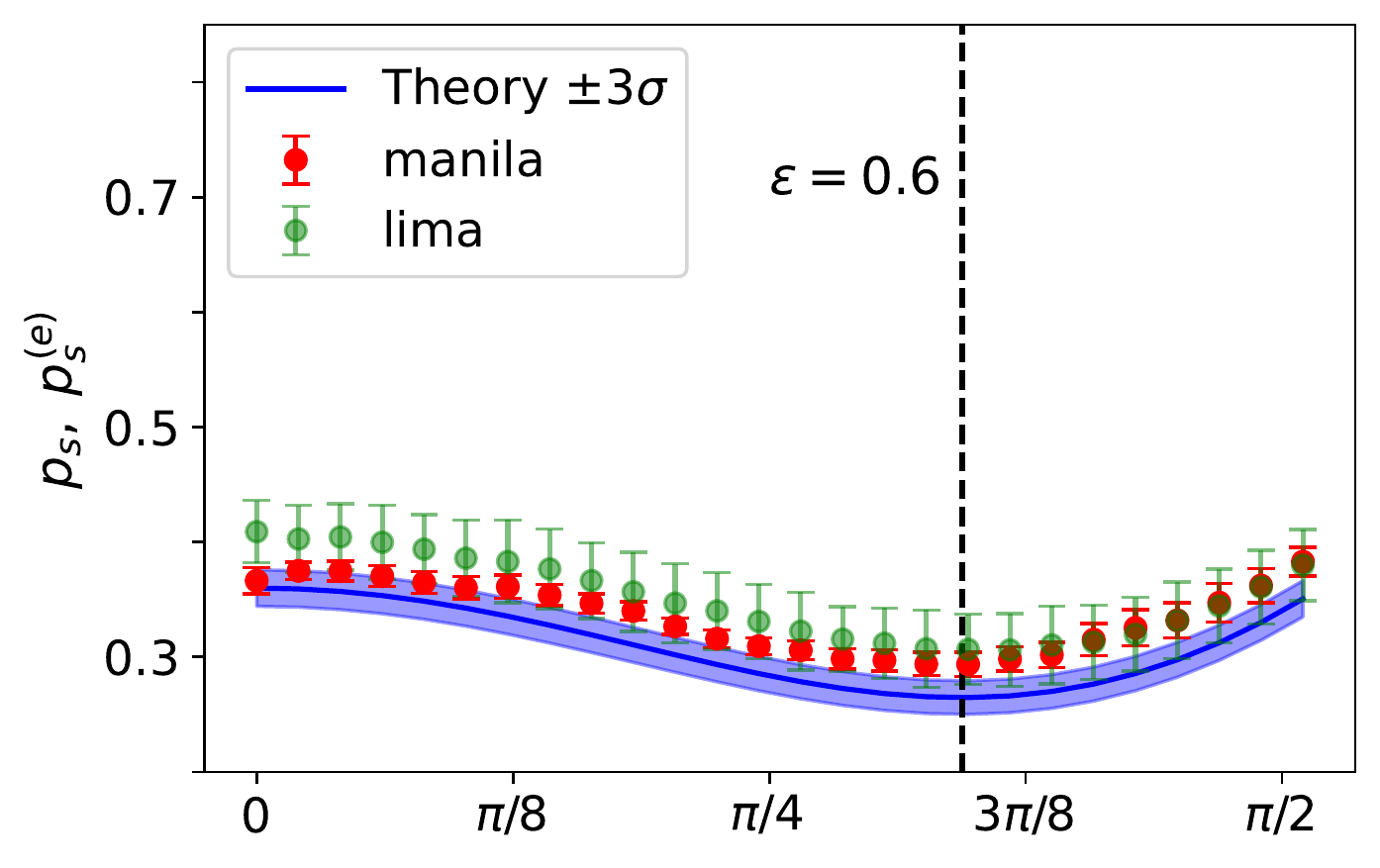}
  \includegraphics[scale=0.55]{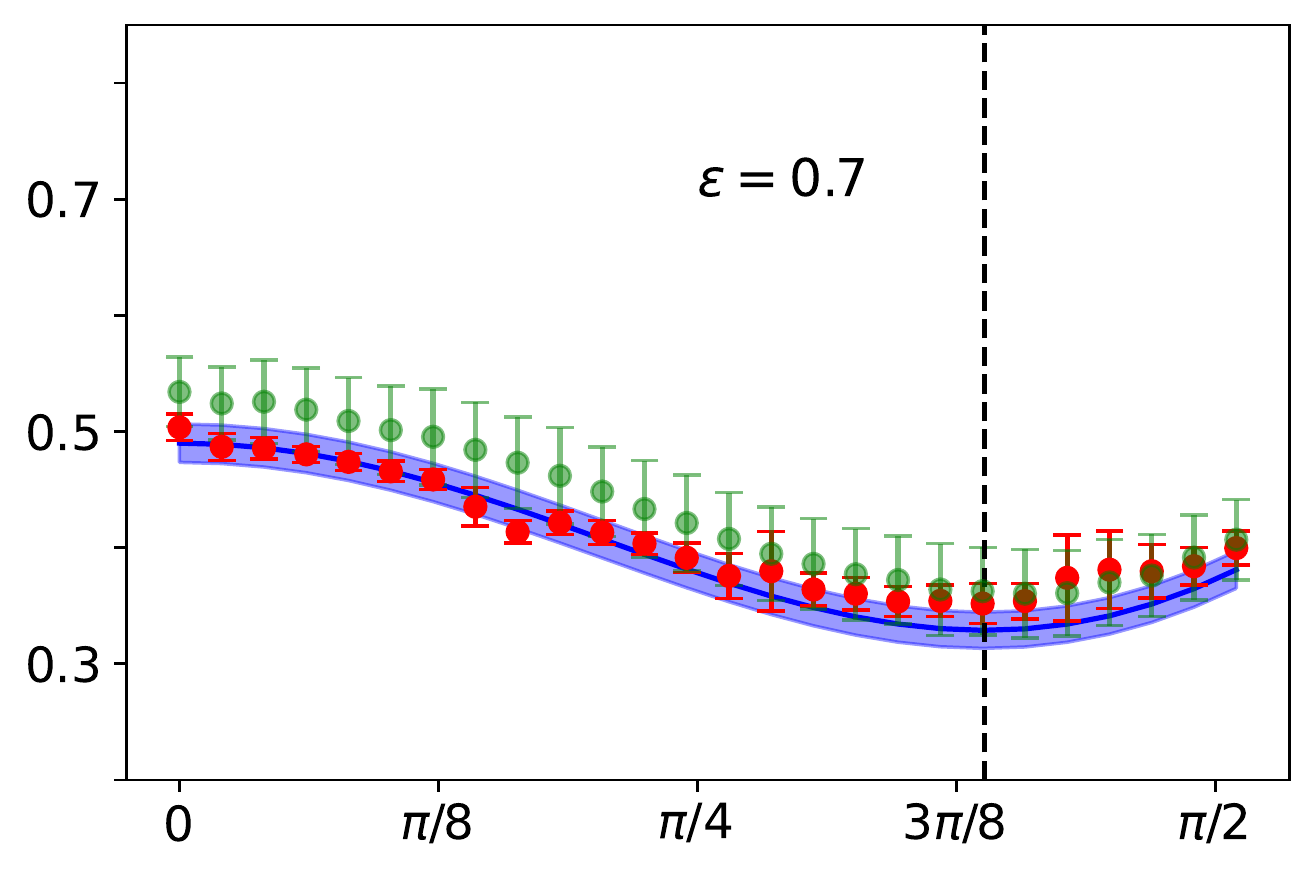}
  \includegraphics[scale=0.55]{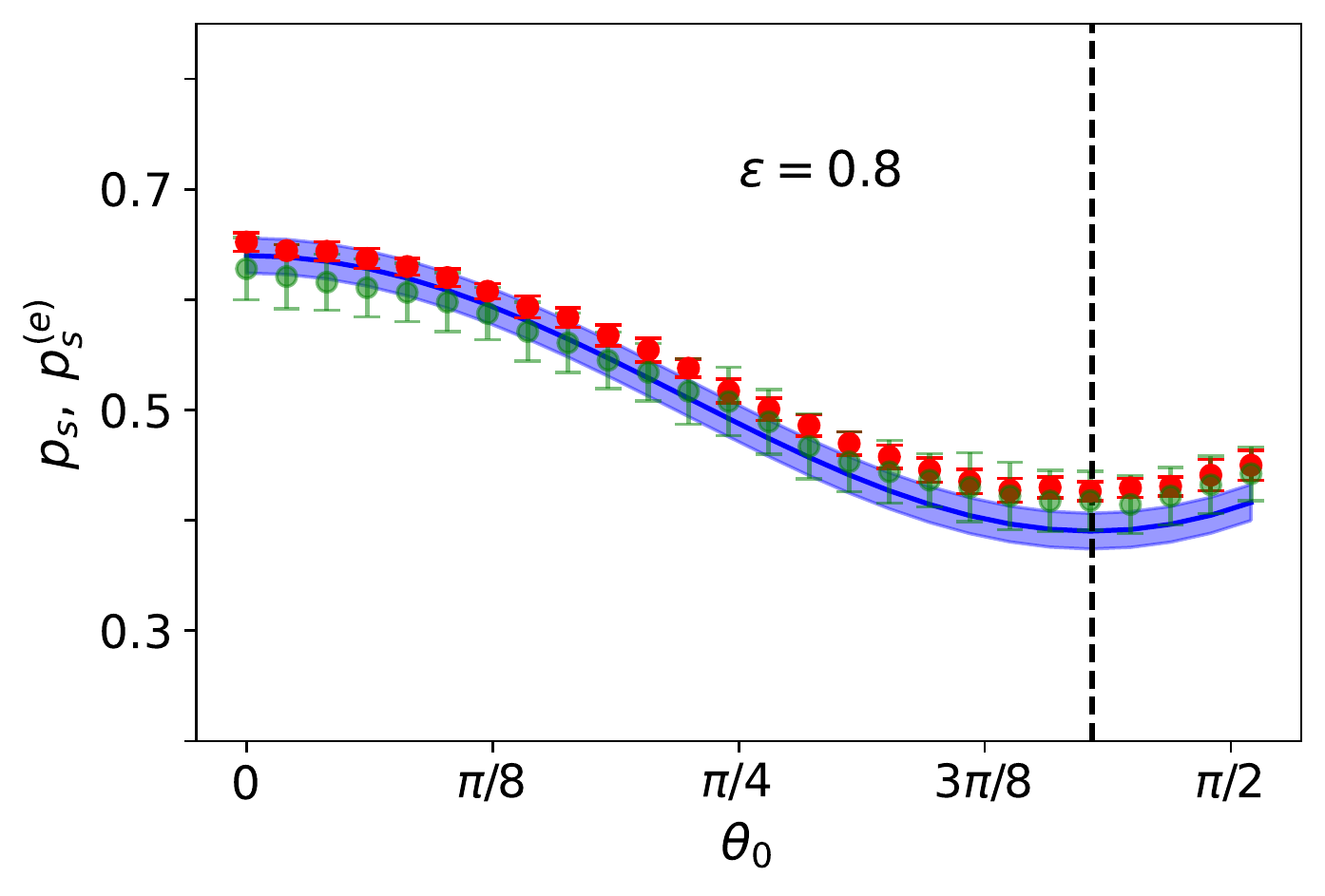}
  \includegraphics[scale=0.55]{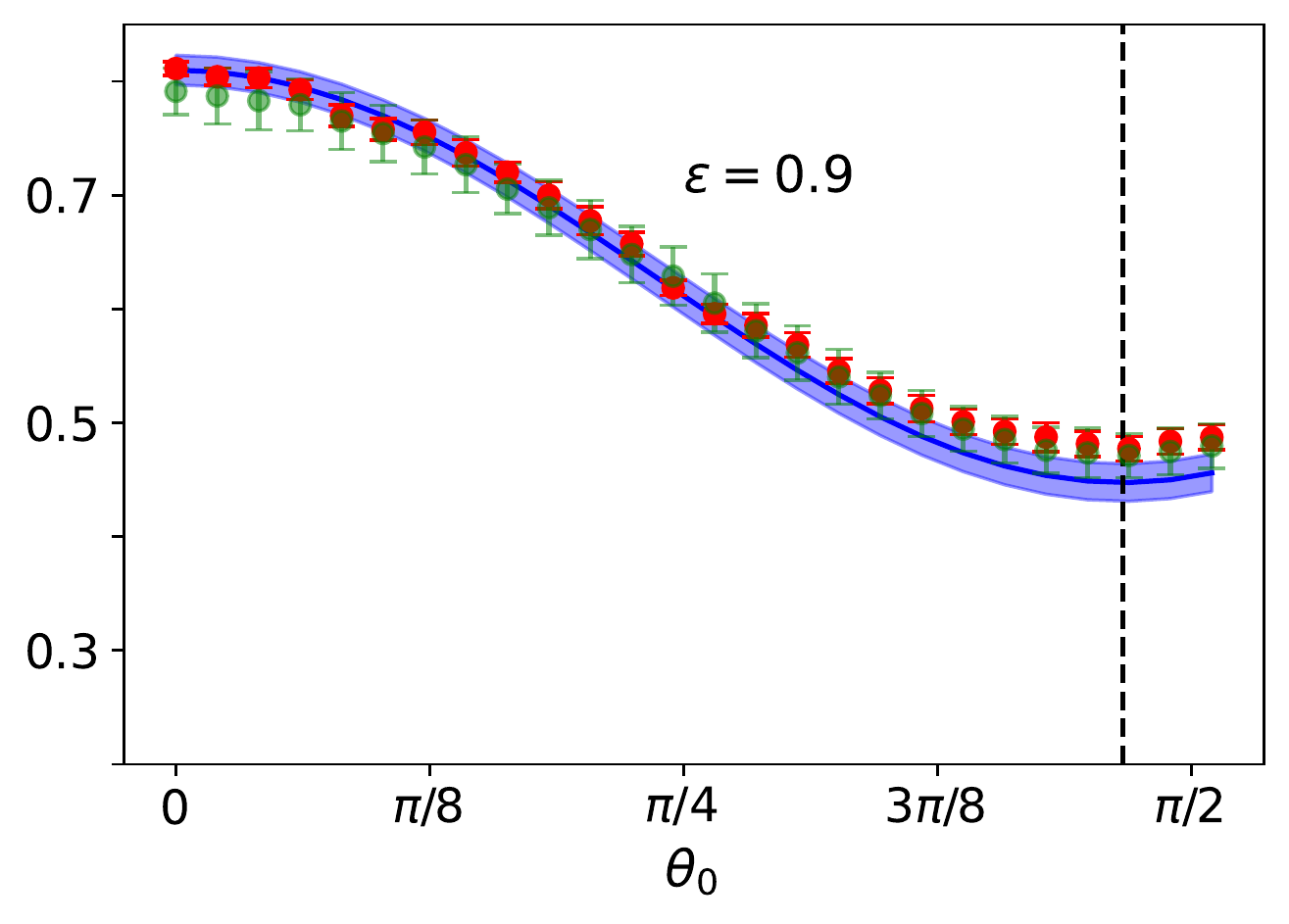}
  
  \caption{(color online) The relative frequency of success $p_{s}^{(e)}$ of the experimentally implemented quantum state matching protocol for different values of $\epsilon$ as a function of $\theta_{0}$. For every value of $\theta_{0}$, $25$ different, equally spaced values of $\phi_{0}$ were taken. Red and green dots represent the average of the experimentally obtained relative frequencies of success $p_{s}^{(e)}$ of \latdev{ibmq\_manila} and \latdev{ibmq\_lima}, respectively. The error bars represent the standard deviations of the results for a given value of $\theta_{0}$. The solid blue line represents the theoretical value of the success probability $p_s$, while the blue shaded area represents $\pm3\sigma$ variance, corresponding to the statistical uncertainty of having finite number of repetitions (in this case $2^{13}$ shots per a given configuration of $(\epsilon, \theta_{0}, \phi_{0})$.}
  \label{fig:res1ps}
\end{figure}

In order to test the devices, we ran the circuit for four different cases of $\epsilon$. In every case, we took $26$ different, equally spaced values of $\theta_0$ in the interval $[0,25\pi/49]$. For every value of $\theta_{0}$ we repeated the experiment $25$ times, $2^{13}$ shots each, with equally spaced values of $\phi_{0}$ in the interval $\left[0,2\pi\right]$. (We note that the state preparation step, i.e., the setting of the values of $\theta_{0}$ and $\phi_{0}$ were achieved by applying an $R_y(\theta_{0})$ gate and a subsequent $P(\phi_{0})$ phase gate.)
The results are shown in Fig.~\ref{fig:res1ps}.

It can be seen in Fig.~\ref{fig:res1ps} that the higher the value of $\epsilon$, the closer $p_{s}^{(e)}$ is to the ideal theoretical success probability in the case of both devices. The size of the error bars also decrease as $\epsilon$ is increased. When we compare the two devices, it can be seen that for lower values of $\epsilon$, the device with the larger QV value (\latdev{ibmq\_manila}) performs better: not only the average value of $p_{s}^{(e)}$ is closer to the theoretical value, but also the standard deviation of the results are smaller. In the case of $\epsilon=0.6$ and $0.7$, all the results of \latdev{ibmq\_lima} fall outside the shaded blue region representing the statistical tolerance. This feature reveals that device errors, clearly distinguishable from statistical errors, are present in this quantum computer. The fact that $p_{s}^{(e)}$ overestimates $p_{s}$ for every $(\theta_{0},\phi_{0})$ in these two cases of $\epsilon$ suggest that the decay of the qubits from the excited to the ground state, especially the transition $|01\rangle \rightarrow |00\rangle$ might affect the results, as these lead to extra counts corresponding to the state $\ket{00}$ even though they do not represent a successful implementation of the protocol.

We note that our results suggest that lower values of $\epsilon$ or a higher number of iterations would result in even larger differences of $p_{s}^{(e)}$ from the theoretical values, indicating that device errors are quite significant in the tested qcs. Yet, we can assess that the technology have shown important improvements compared to previous analysis~\cite{MichielsenCPC2017}, where most of the time ``raw" experiments did not yield successful results.

We can also investigate the performance of the devices by reconstructing the transformed quantum state using the $p^{(e)}_{00}$ and $p^{(e)}_{10}$ relative frequencies of the measurement results '00' and '10', respectively. The angle $\theta_1$ after one step is then estimated by

\begin{equation}
    \theta_1^{(e)} = 2\,  \arctan\left(\sqrt{\frac{p^{(e)}_{10}}{p^{(e)}_{00}}}\right).
    \label{eq:theta1}
\end{equation}

\begin{figure}[htb]
  \centering
  \hspace{-2cm}
  \includegraphics[scale=0.6]{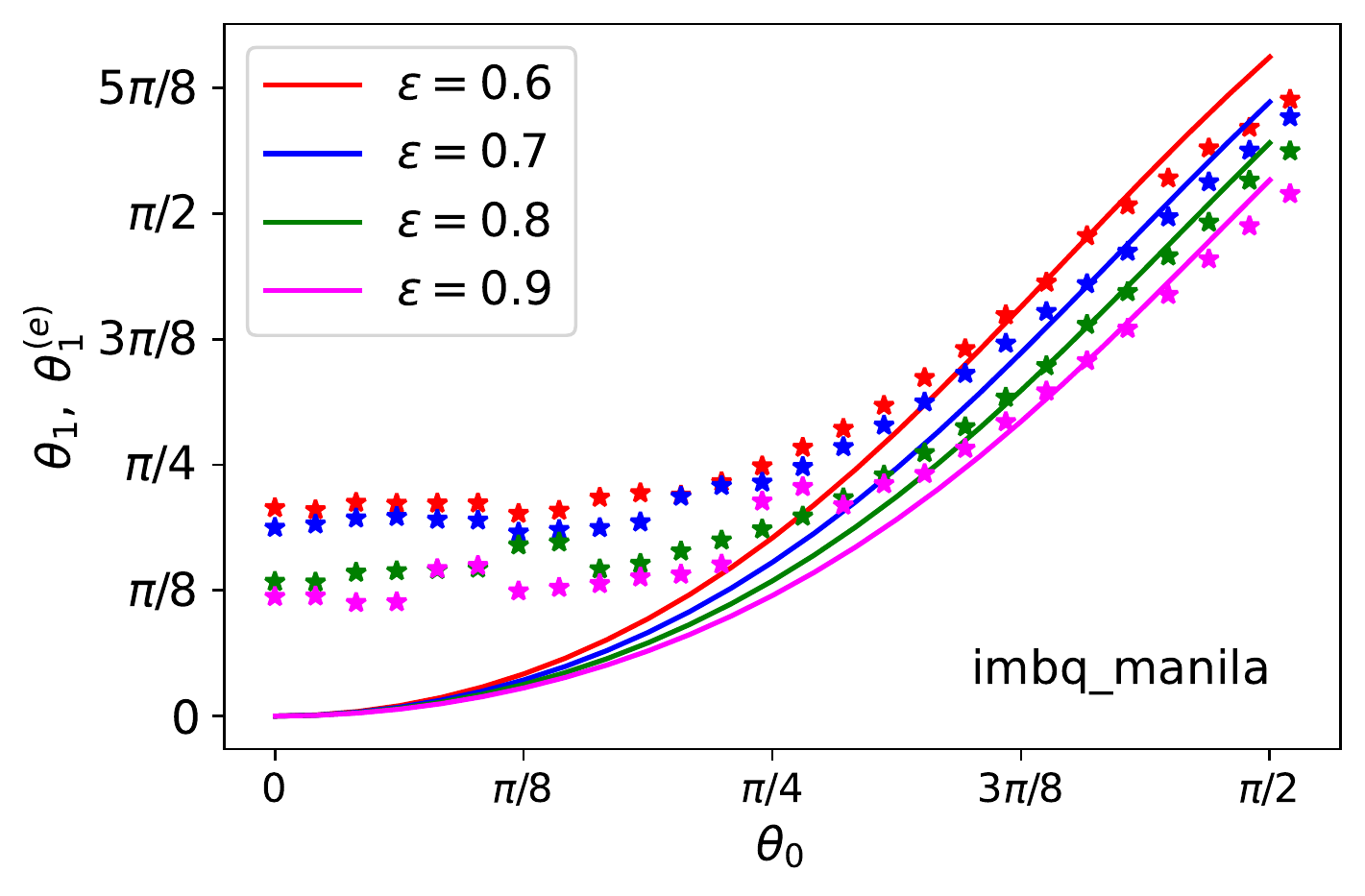}
  \includegraphics[scale=0.6]{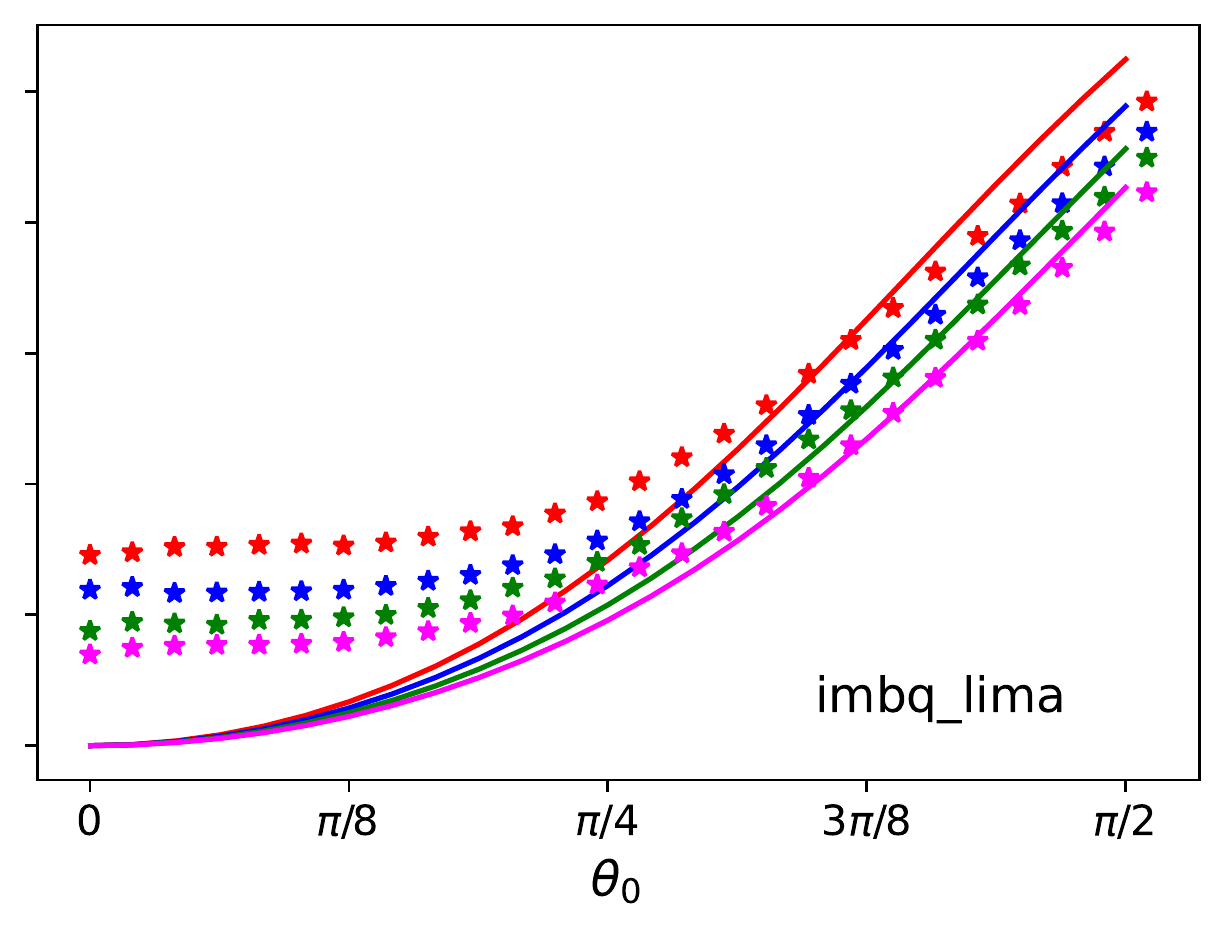}
  \caption{(color online) The theoretical angle $\theta_1$ (solid lines) and the experimentally determined angle $\theta_{1}^{(e)}$ (stars) after one step of the protocol as a function of the initial angle $\theta_0$. (left) Results from \latdev{ibmq\_manila}, (right) results from \latdev{ibmq\_lima}. The experiments were realized for the same input parameters as in Fig.~\ref{fig:res1ps}. Stars denote the average values of $\theta_{1}^{(e)}$.}
  \label{fig:res1theta1}
\end{figure}

 Figure~\ref{fig:res1theta1} shows $\theta_1$ and $\theta_{1}^{(e)}$ as a function of the initial $\theta_0$. One can see that the results of both devices deviate from the ideal value for most inputs. For lower values of $\theta_{0}$ ($\theta_{0}\lessapprox \pi/4$) the angle $\theta_{1}$ is generally overestimated. This artifact may be related to measuring extra counts in '10', resulting from e.g. a $\ket{11}$ $\rightarrow$ $\ket{10}$ decay process. We will show in Sec.~\ref{sec:errmit} that some part of these errors can be mitigated by applying a readout error mitigation scheme. The remaining deviations may be caused by other processes (such as imperfect state preparation, and gate operations), which we do not attempt to model in this work. 

 A striking difference between the two devices is that, while in. Fig.~\ref{fig:res1ps} \latdev{ibmq\_manila} seems to perform better in terms of the estimated success probability, when estimating the quantum state itself (with the angle $\theta_1^{(e)}$), then the results from \latdev{ibmq\_lima} are more regular than those of \latdev{ibmq\_manila}. This indicates that \latdev{ibmq\_lima} is more stable in terms of the overall operations, as well as over time than \latdev{ibmq\_manila}.
 
\subsection{Randomly chosen $\phi_{0}$ inputs}
\label{sec:randinput}

As we have mentioned in Sec.~\ref{sec:fram}, the theoretical success probability $p_{s}$ is independent of $\phi_0$ (for a fixed $\epsilon$ and $\theta_0$). In an actual implementation on a qc there may be errors at any part of the circuit during its execution, which may lead to a $p_{s}^{(e)}$ that is depending on $\phi_{0}$ or fluctuating as $\phi_{0}$ is changed.

\begin{figure}[htb]
  \centering
  \includegraphics[scale=0.55]{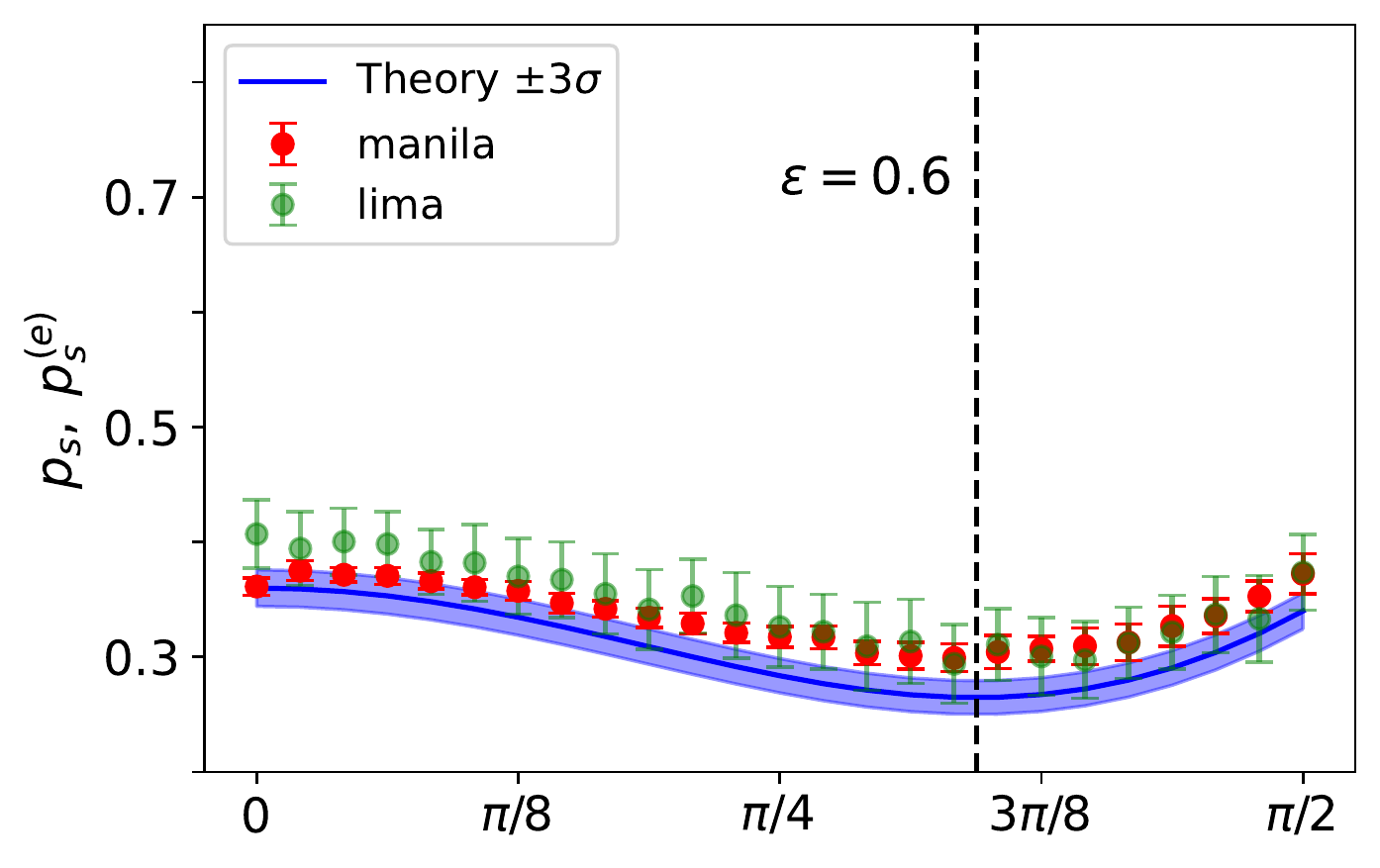}
  \includegraphics[scale=0.55]{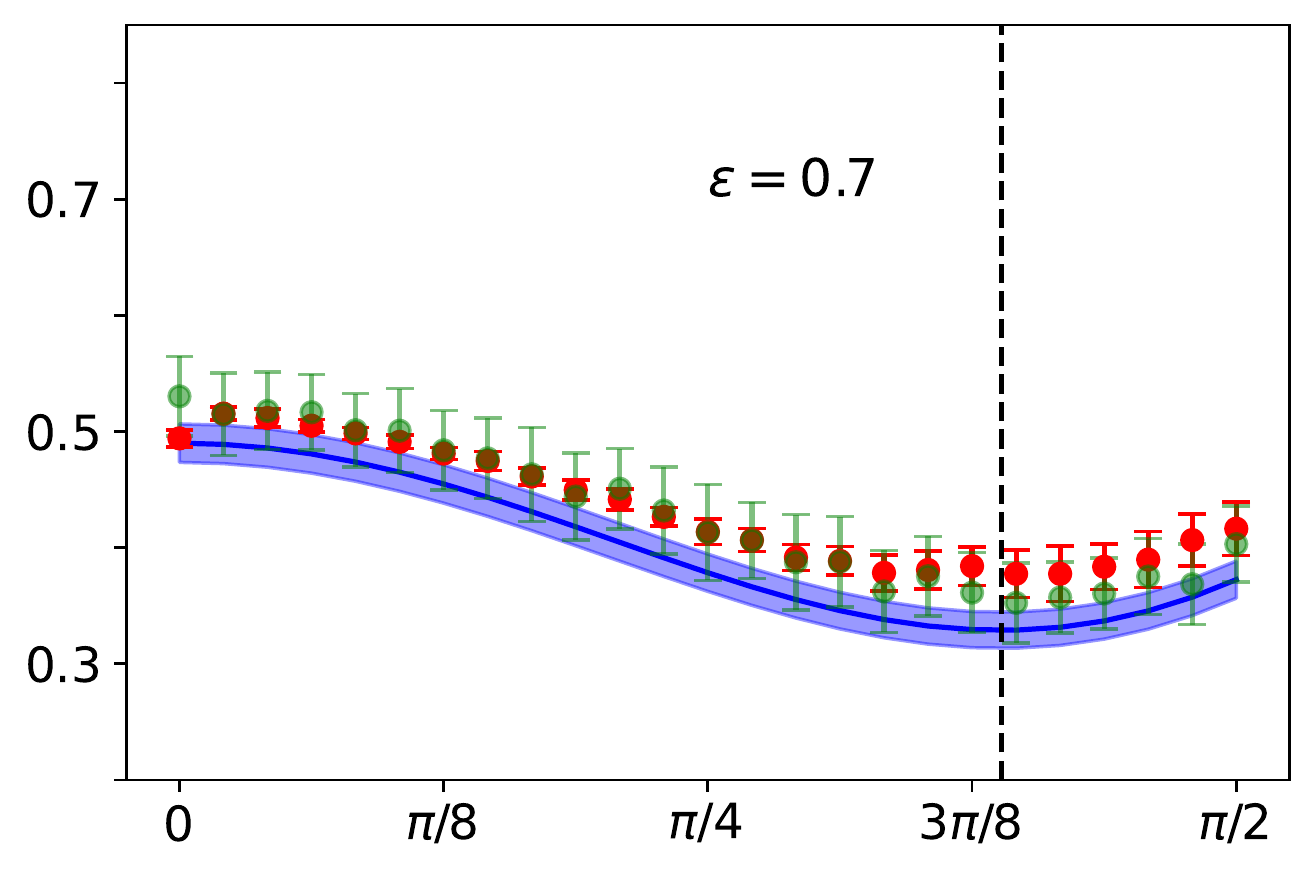}
  \includegraphics[scale=0.55]{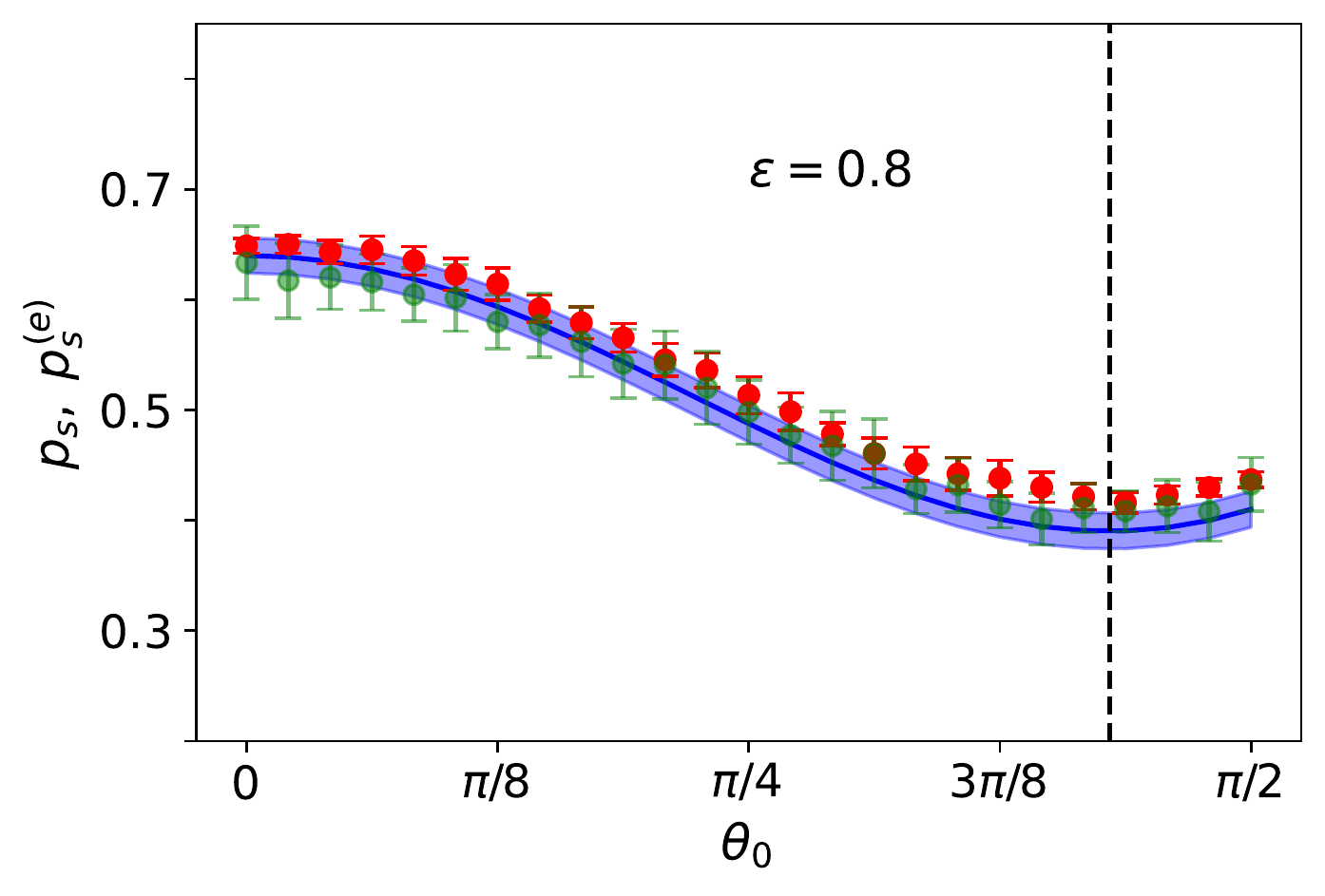}
  \includegraphics[scale=0.55]{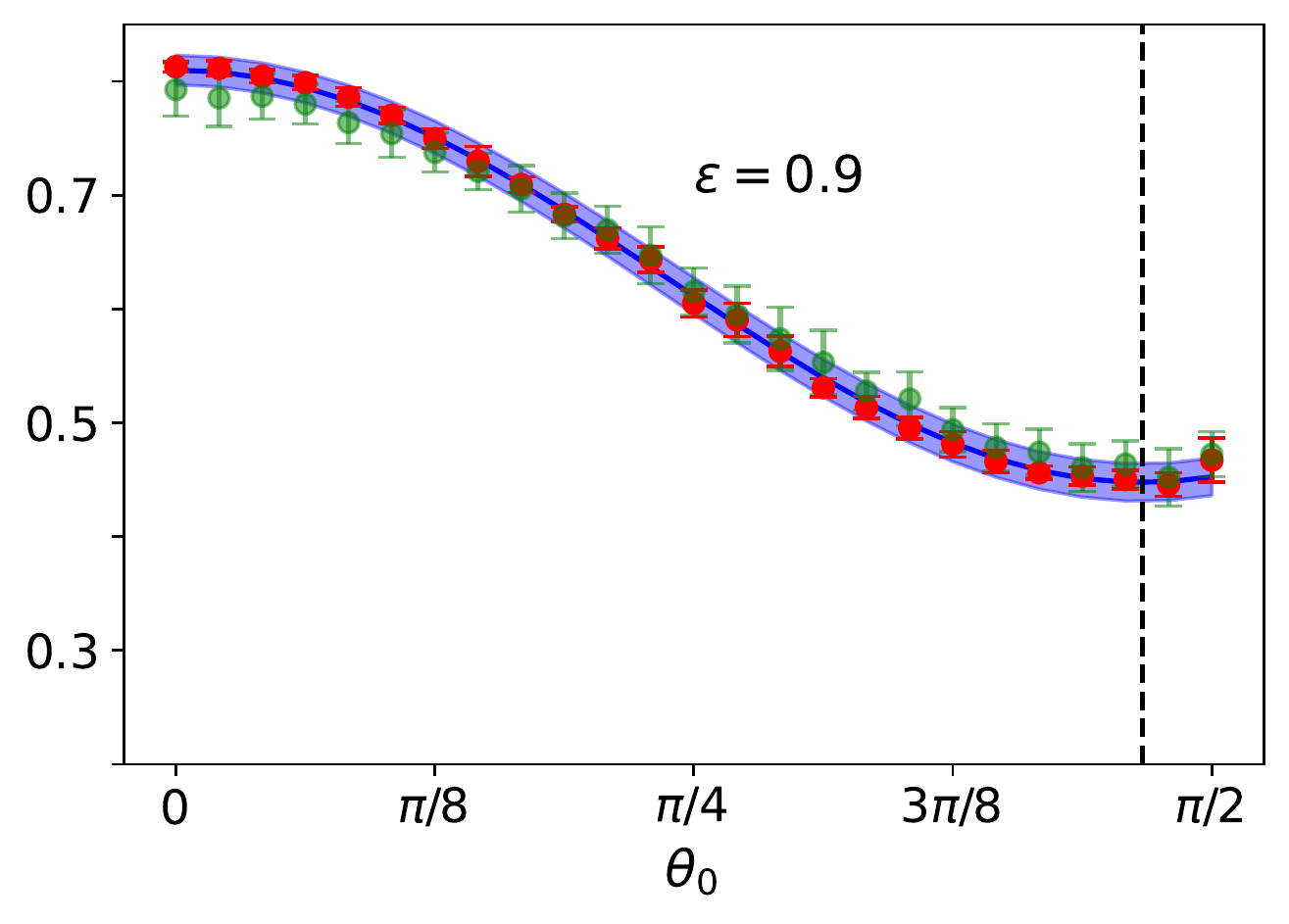}
  
  \caption{(color online) Relative frequency of success $p_{s}^{(e)}$ and theoretical success probability $p_s$ for four different values of $\epsilon$ as a function of $\theta_{0}$. For every value of $\theta_{0}$, 25 randomly chosen values of $\phi_{0}$ were taken. Notations are similar to those of Fig.~\ref{fig:res1ps}.}
  \label{fig:res2ps}
\end{figure}

In Figs.~\ref{fig:res1ps} and \ref{fig:res1theta1}, for every fixed value of $\theta_0$ we used the same 25 equally spaced values of $\phi_0$, which gives a fairly regular ensemble of inputs. Another way to test a device is by altering the way we choose the input values of $\phi_{0}$. Figure~\ref{fig:res2ps} shows the results  obtained by choosing for every $\theta_0$ a random sequence of 25 $\phi_0$-values.

This figure can be considered as a complement of Fig.~\ref{fig:res1ps}: some features are similar in the overall picture (e.g., lower values of $\epsilon$ result in weaker performance), but we also find some differences arising from using a random ensemble of inputs. For instance, if one looks at the $\epsilon=0.6$ and $0.7$ cases, it can be seen that the $p_{s}^{(e)}$ values for \latdev{ibmq\_lima} are closer to the results
of \latdev{ibmq\_manila}, albeit the same random values of $\phi_0$ were used for both devices. This suggests that using a random set of inputs might conceal some errors. Consequently, the results in a NISQ system can be heavily influenced by choosing appropriate sets of parameters for a given quantum circuit: some can reveal an optimistic performance (as suggested by the random inputs and the QV values), or they can reveal device errors (as suggested by our results using systematic inputs and gate sets).

\subsection{Readout error mitigation}
\label{sec:errmit}

One of the problems of NISQ systems is the reduced number of qubits available. In order to implement a universal fault-tolerant quantum computer, one would need about $10^6$ physical qubits with low error rates and long coherence times~\cite{AspuruRMP2022}. At the time of writing this paper, the largest IBM quantum device is the \latdev{ibmq\_washington} with 127 qubits. It is clear that in the near future fault-tolerant quantum computers will not be available. If one is interested in reliable results from NISQ quantum computers, other frameworks of quantum correction must be utilized. It is known that in superconducting qubits a significant amount of errors come from the detection step \cite{ZimborasQuantum2022}. One way to reduce such errors is to use a readout error mitigation procedure\cite{QiskitTextbook,ZimborasQuantum2022}. In this section, we apply such a scheme to our results. Without loss of generality, we only provide the analysis for the case of randomly chosen initial values of $\phi_{0}$, corresponding to the results shown in Fig.~\ref{fig:res2ps}.

\begin{figure}[H]
  \centering
  \includegraphics[width=0.95\textwidth]{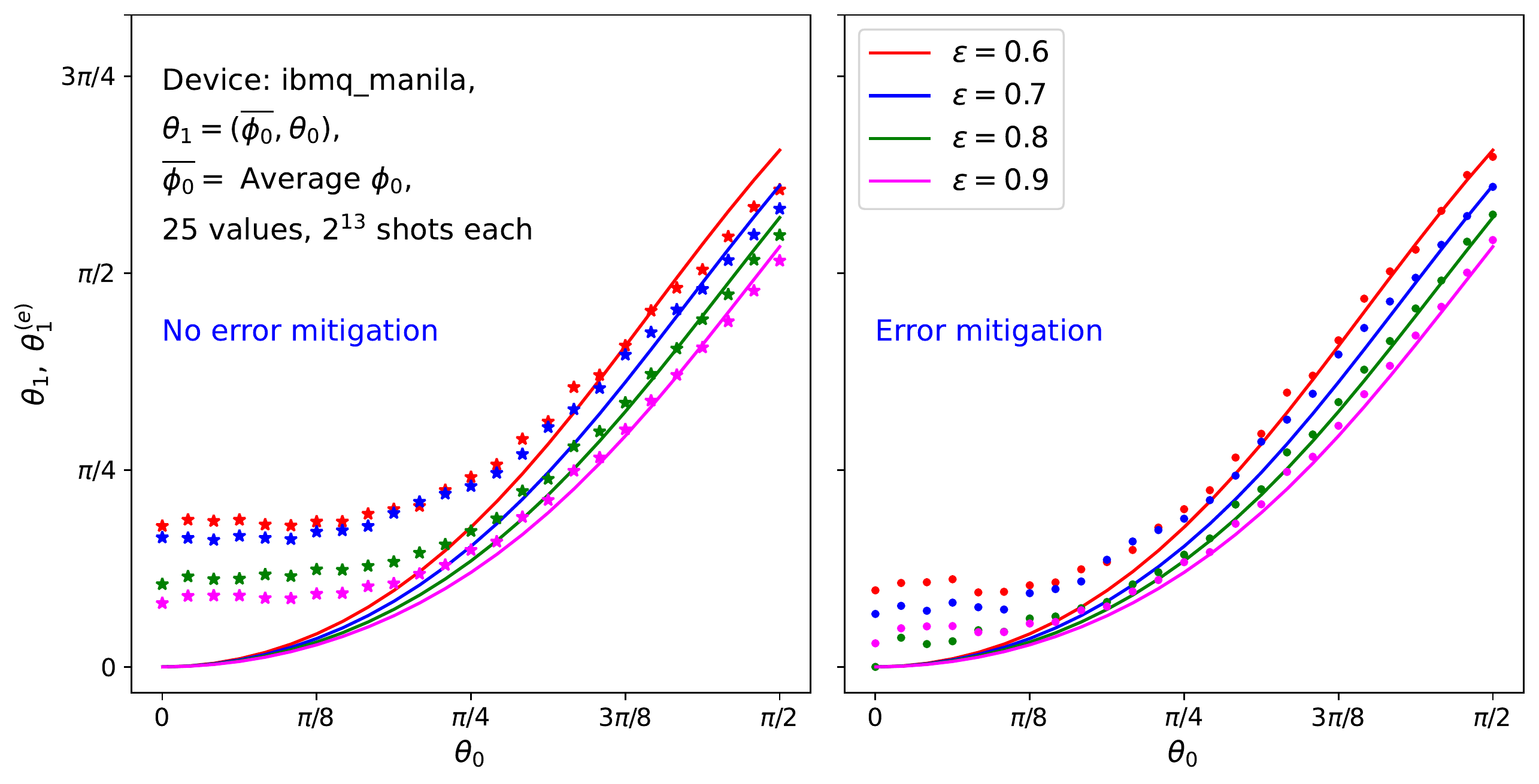}
  \includegraphics[width=0.95\textwidth]{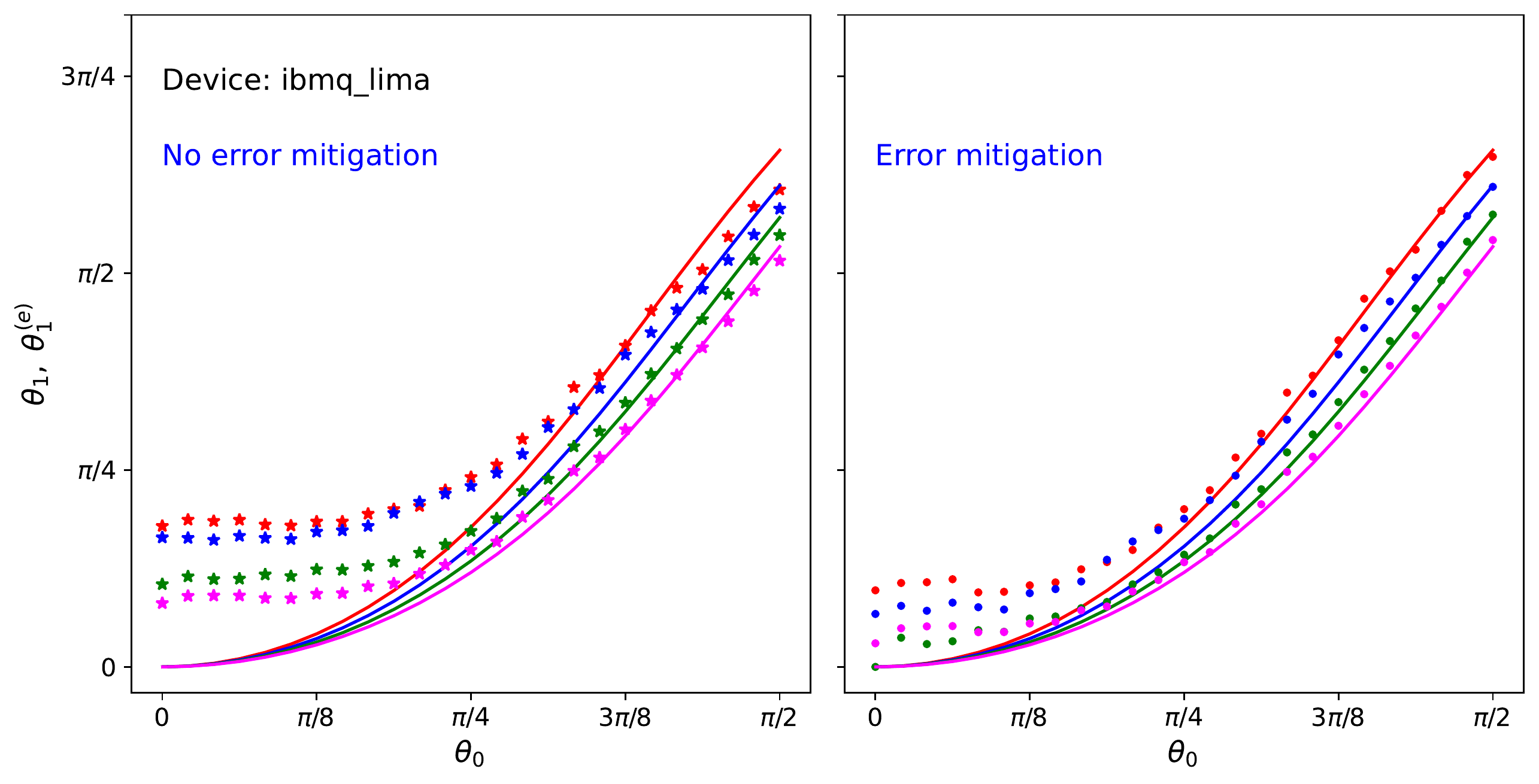}
  \caption{(color online) angle $\theta_1$ (solid lines) and $\theta_{1}^{(e)}$ (stars, dots) after one step of the protocol as a function of the initial angle $\theta_0$ before (left) and after applying readout error mitigation (right). Results from \latdev{ibmq\_manila} (\latdev{ibmq\_lima}) are displayed in the top (bottom) row. The results were obtained from the same data as those in Fig.~\ref{fig:res2ps}.} 
  \label{fig:res2}
\end{figure}

Here we focus on the impact of readout error mitigation on the experimentally estimated values of $\theta_1$. 
Figure~\ref{fig:res2} shows the results without (left column) and with (right column) error mitigation, as a function of the initial angle $\theta_0$. The mitigation matrices~\cite{QiskitTextbook} used in the procedure were determined directly by carrying out the full tomography of the two qubits used in the circuit. It can be seen that the readout error mitigation procedure can significantly improve the results, even though the tomography could not be done at the same time as the experiment. Our results are consistent with the assumption that the readout step in these devices is indeed quite erroneous.

\section{Summary and outlook}
\label{sec:Conclusions}

We have presented implementations of the quantum state matching protocol on two IBM quantum computers based on superconducting qubits. The most important ingredient of the protocol is a specific two-qubit entangling unitary $U_{\epsilon}$. We have determined the optimal decomposition of $U_{\epsilon}$ with programmable quantum gates, and implemented the quantum circuit corresponding to one step of the protocol for systematically varied and randomly chosen input parameters as well. 

We found a qualitative disagreement between the Quantum Volume of the devices and our results, obtained by the implementation of the quantum state matching protocol. We also showed that randomly chosen inputs may change the results and give the impression of a better performance. Further, we have also implemented a readout error mitigation procedure on our results which removed some of the errors, suggesting that detection errors are indeed quite significant in these superconducting qcs. 

In our analysis, we determined the relative frequency of success experimentally and compared it to the ideal theoretical success probability with $\pm 3\sigma$ statistical tolerances to test how well a quantum circuit is implemented in current qcs. We showed that already in the simplest cases of our protocol, and with this very simple quantity, the tested devices behaved opposite to what one would have expected from their Quantum Volumes~\cite{CrossPRA2019,BaldwinQu2019}. This suggests that our test is more demanding than that used for the calculation of the QV. The success probability in our protocol can be tuned to  any value, while in the Quantum Volume framework, the ``success probability" always includes all the heavy outputs coming from the full probability distribution of the final state. 

The most important advantage of our protocol in testing quantum computers is that it is scalable: irrespective of the number of steps, the expected result can be classically calculated, despite the fact that, in a qc realization, the protocol requires $2^n$ qubits for $n$ steps. In a future work we shall test more iterations of our protocol on larger devices, based on several different physical systems as well (including IonQ or Rigetti).

\section*{Acknowledgements}
We are grateful for the support of the National Research, Development and Innovation Office of Hungary (Project No. K124351) and the Quantum Information National Laboratory of Hungary.
We are grateful to Igor Jex for the long-term collaboration exploring among others nonlinear quantum dynamics, quantum walks and other interesting aspects of life.
\appendix
\section{Dates of the experiments}
\label{sec:app2}

In this appendix we provide the dates of the experiments run on the two devices. 
The format is year-month-day-(days) and the order is in increasing value of $\epsilon$ according to the referenced figures.

\begin{itemize}
    \item Figure~\ref{fig:res1ps}, \ref{fig:res1theta1}, \latdev{ibmq\_manila}:  2022-08-04-05,  2021-06-25, 2022-08-08, 2022-08-09; \latdev{ibmq\_lima}: 2022-08-04-08, 2022-08-03-04, 2022-08-08, 2022-08-09. 
    \item Figure~\ref{fig:res2ps}, \ref{fig:res2}, \latdev{ibmq\_manila}: 2021-06-22, 2021-06-23, 2021-06-23, 2021-06-11; \latdev{ibmq\_lima}: 2022-07-29, 2022-08-01, 2022-08-01, 2022-08-02. 
\end{itemize}

\bibliographystyle{hunsrt}
\bibliography{bibliography.bib}

\begin{thebibliography}{10}

\bibitem{IBMqcs}
{IBM Quantum}.
\newblock \url{ https://quantum-computing.ibm.com/}.
\newblock Accessed: 2022-06.

\bibitem{rigettiqc}
{Rigetti}.
\newblock \url{https://www.rigetti.com/}.
\newblock Accessed: 2022-06.

\bibitem{oxfordqc}
{Oxford}.
\newblock \url{https://oxfordquantumcircuits.com/oqc-on-aws}.
\newblock Accessed: 2022-06.

\bibitem{ionqqc}
{IonQ}.
\newblock \url{https://ionq.com/}.
\newblock Accessed: 2022-06.

\bibitem{quixqc}
{Quix}.
\newblock \url{https://www.quixquantum.com/}.
\newblock Accessed: 2022-06.

\bibitem{pasqalqc}
{Pasqal}.
\newblock \url{https://pasqal.io/}.
\newblock Accessed: 2022-06.

\bibitem{AspuruRMP2022}
Kishor Bharti, Alba Cervera-Lierta, Thi~Ha Kyaw, Tobias Haug, Sumner
  Alperin-Lea, Abhinav Anand, Matthias Degroote, Hermanni Heimonen, Jakob~S.
  Kottmann, Tim Menke, Wai-Keong Mok, Sukin Sim, Leong-Chuan Kwek, and Al\'an
  Aspuru-Guzik.
\newblock Noisy intermediate-scale quantum algorithms.
\newblock {\em Rev. Mod. Phys.}, 94:015004, Feb 2022.

\bibitem{Hennessybook2017}
J.~L. Hennessy and D.~A. Patterson.
\newblock {\em Computer Architecture: A Quantitative Approach}.
\newblock Morgan Kaufmann, 6 edition, 2017.

\bibitem{BlumeOSTI2019}
Robin~J Blume-Kohout and Kevin Young.
\newblock Metrics and benchmarks for quantum processors: State of play.
\newblock 1 2019.

\bibitem{MollQST2018}
Nikolaj Moll, Panagiotis Barkoutsos, Lev~S Bishop, Jerry~M Chow, Andrew Cross,
  Daniel~J Egger, Stefan Filipp, Andreas Fuhrer, Jay~M Gambetta, Marc Ganzhorn,
  Abhinav Kandala, Antonio Mezzacapo, Peter Müller, Walter Riess, Gian Salis,
  John Smolin, Ivano Tavernelli, and Kristan Temme.
\newblock Quantum optimization using variational algorithms on near-term
  quantum devices.
\newblock {\em Quantum Science and Technology}, 3(3):030503, jun 2018.

\bibitem{CrossPRA2019}
Andrew~W. Cross, Lev~S. Bishop, Sarah Sheldon, Paul~D. Nation, and Jay~M.
  Gambetta.
\newblock Validating quantum computers using randomized model circuits.
\newblock {\em Phys. Rev. A}, 100:032328, Sep 2019.

\bibitem{atosqscore}
{$Q$-score}.
\newblock \url{https://atos.net/en/solutions/q-score}.
\newblock Accessed: 2022-06.

\bibitem{DasguptaArxiv2020}
Samudra Dasgupta and Travis~S. Humble.
\newblock Characterizing the stability of {NISQ} devices, 2020,
  arXiv:quant-ph/2008.09612.

\bibitem{EmersonJOptB2005}
Joseph Emerson, Robert Alicki, and Karol Życzkowski.
\newblock Scalable noise estimation with random unitary operators.
\newblock {\em Journal of Optics B: Quantum and Semiclassical Optics},
  7(10):S347, 2005.

\bibitem{KnillPRA2008}
E.~Knill, D.~Leibfried, R.~Reichle, J.~Britton, R.~B. Blakestad, J.~D. Jost,
  C.~Langer, R.~Ozeri, S.~Seidelin, and D.~J. Wineland.
\newblock Randomized benchmarking of quantum gates.
\newblock {\em Phys. Rev. A}, 77:012307, 2008.

\bibitem{MagesanPRL2011}
Easwar Magesan, J.~M. Gambetta, and Joseph Emerson.
\newblock Scalable and robust randomized benchmarking of quantum processes.
\newblock {\em Phys. Rev. Lett.}, 106:180504, 2011.

\bibitem{MagesanPRA2012}
Easwar Magesan, Jay~M. Gambetta, and Joseph Emerson.
\newblock Characterizing quantum gates via randomized benchmarking.
\newblock {\em Phys. Rev. A}, 85:042311, 2012.

\bibitem{ProctorNatPhys2021}
T.~Proctor, K.~Rudinger, K.~Young, E.~Nielsen, and R.~Blume-Kohout.
\newblock Measuring the capabilities of quantum computers.
\newblock {\em Nat. Phys.}, 18:75, 2021.

\bibitem{LinkePNAS2017}
N.~M. Linke, D~Maslovc, M.~Roettelerd, S.~Debnath, C.~Figgatt, K.~A. Landsman,
  K.~Wright, and C.~Monroe.
\newblock Experimental comparison of two quantumcomputing architectures.
\newblock {\em PNAS}, 114:3305, 2017.

\bibitem{WrightNatCom2019}
K.~Wright, K.~M. Beck, S.~Debnath, J.~M. Amini, Y.~Nam, N.~Grzesiak, J.-S.
  Chen, N.~C. Pisenti, M.~Chmielewski, C.~Collins, K.~M. Hudek, J.~Mizrahi,
  J.~D. Wong-Campos, S.~Allen, J.~Apisdorf, P.~Solomon, M.~Williams, A.~M.
  Ducore, A.~Blinov, S.~M. Kreikemeier, V.~Chaplin, M.~Keesan, C.~Monroe, and
  J.~Kim.
\newblock Benchmarking an 11-qubit quantum computer.
\newblock {\em Nature {C}ommunications}, 10:5464, 2019.

\bibitem{ZimborasQuantum2022}
F~B Maciejewski, Z~Zimbor\'as, and M~Oszmaniec.
\newblock Mitigation of readout noise in near-term quantum devices by classical
  post-processing based on detector tomography.
\newblock {\em Quantum}, 6:707, 2022.

\bibitem{GilyenArxiv2021}
Arjan Cornelissen, Johannes Bausch, and András Gilyén.
\newblock Scalable benchmarks for gate-based quantum computers, 2021,
  arXiv:2104.10698.

\bibitem{McCaskeyNQI2019}
A.~J. McCaskey, Z.~P. Parks, J.~Jakowski, S.~V. Moore, T.~D. Morris, T.~Humble,
  and R.~C. Pooser.
\newblock Quantum chemistry as a benchmark for near-term quantum computers.
\newblock {\em npj Quantum Inf.}, 5:99, 2019.

\bibitem{KalmanPRA2018}
Orsolya K\'alm\'an and Tam\'as Kiss.
\newblock Quantum state matching of qubits via measurement-induced nonlinear
  transformations.
\newblock {\em Phys. Rev. A}, 97:032125, Mar 2018.

\bibitem{Milnorbook2011}
J.~Milnor.
\newblock {\em Dynamics in One Complex Variable}.
\newblock Princeton University Press, 3 edition, 2011.

\bibitem{KalmanJRLR2018}
O.~K\'{a}lm\'{a}n, T.~Kiss, and I.~Jex.
\newblock Sensitivity to initial noise in measurement-induced nonlinear quantum
  dynamics.
\newblock {\em J. Russ. Laser Res.}, 39:382, 2018.

\bibitem{Beardonbook1991}
A.~F. Beardon.
\newblock {\em Iteration of Rational Functions: Complex Analytic Dynamical
  Systems}.
\newblock Springer, 1 edition, 1991.

\bibitem{KrausPRA2001}
B.~Kraus and J.~I. Cirac.
\newblock Optimal creation of entanglement using a two-qubit gate.
\newblock {\em Phys. Rev. A}, 63:062309, May 2001.

\bibitem{VidalPRA2004}
G.~Vidal and C.~M. Dawson.
\newblock Universal quantum circuit for two-qubit transformations with three
  controlled-not gates.
\newblock {\em Phys. Rev. A}, 69:010301, Jan 2004.

\bibitem{TucciArxiv2005}
Robert~R. Tucci.
\newblock An introduction to {C}artan's {KAK} decomposition for {QC}
  programmers, 2005, arXiv:quant-ph/0507171.

\bibitem{BullockPRA2003}
Stephen~S. Bullock and Igor~L. Markov.
\newblock Arbitrary two-qubit computation in 23 elementary gates.
\newblock {\em Phys. Rev. A}, 68:012318, Jul 2003.

\bibitem{VatanPRA2004}
Farrokh Vatan and Colin Williams.
\newblock Optimal quantum circuits for general two-qubit gates.
\newblock {\em Phys. Rev. A}, 69:032315, Mar 2004.

\bibitem{KhanejaArxiv2000}
Navin Khaneja and Steffen Glaser.
\newblock Cartan decomposition of {$SU(2^n)$}, constructive controllability of
  spin systems and universal quantum computing, 2000, quant-ph/0010100.

\bibitem{KhanejaPRA2001}
N.~Khaneja, R.~Brockett, and S.~J. Glaser.
\newblock Time optimal control in spin systems.
\newblock {\em Phys. Rev. A}, 63:032308, 2001.

\bibitem{EckartBAMS1939}
C.~Eckart and G.~Young.
\newblock A principal axis transformation for non-hermitian matrices.
\newblock {\em Bull. Amer. Math. Soc.}, 45:118, 1939.

\bibitem{SalonikArxiv2019}
Salonik Resch and Ulya~R. Karpuzcu.
\newblock Benchmarking quantum computers and the impact of quantum noise, 2019,
  arXiv:1912.00546.

\bibitem{MakhlinRMP2001}
Yuriy Makhlin, Gerd Sch\"on, and Alexander Shnirman.
\newblock Quantum-state engineering with josephson-junction devices.
\newblock {\em Rev. Mod. Phys.}, 73:357--400, May 2001.

\bibitem{ChenPRA2019}
Yanzhu Chen, Maziar Farahzad, Shinjae Yoo, and Tzu-Chieh Wei.
\newblock Detector tomography on ibm quantum computers and mitigation of an
  imperfect measurement.
\newblock {\em Phys. Rev. A}, 100:052315, Nov 2019.

\bibitem{AlexandrouArxiv2021}
Constantia Alexandrou, Lena Funcke, Tobias Hartung, Karl Jansen, Stefan Kühn,
  Georgios Polykratis, Paolo Stornati, Xiaoyang Wang, and Tom Weber.
\newblock Investigating the variance increase of readout error mitigation
  through classical bit-flip correction on ibm and rigetti quantum computers,
  2021, arXiv:quant-ph/2111.05026.

\bibitem{ParisLNP2004}
M~Paris and J~\v{R}eh\'{a}\v{c}ek, editors.
\newblock {\em Quantum State Estimation}, volume 649 of {\em Lecture Notes in
  Physics}. Springer, 2004.

\bibitem{MichielsenCPC2017}
Kristel Michielsen, Madita Nocon, Dennis Willsch, Fengping Jin, Thomas Lippert,
  and Hans {De Raedt}.
\newblock Benchmarking gate-based quantum computers.
\newblock {\em Computer Physics Communications}, 220:44--55, 2017.

\bibitem{QiskitTextbook}
Amira Abbas, Stina Andersson, Abraham Asfaw, Antonio Corcoles, Luciano Bello,
  Yael Ben-Haim, Mehdi Bozzo-Rey, Sergey Bravyi, Nicholas Bronn, Lauren
  Capelluto, Almudena~Carrera Vazquez, Jack Ceroni, Richard Chen, Albert
  Frisch, Jay Gambetta, Shelly Garion, Leron Gil, Salvador De La~Puente
  Gonzalez, Francis Harkins, Takashi Imamichi, Pavan Jayasinha, Hwajung Kang,
  Amir h.~Karamlou, Robert Loredo, David McKay, Alberto Maldonado, Antonio
  Macaluso, Antonio Mezzacapo, Zlatko Minev, Ramis Movassagh, Giacomo
  Nannicini, Paul Nation, Anna Phan, Marco Pistoia, Arthur Rattew, Joachim
  Schaefer, Javad Shabani, John Smolin, John Stenger, Kristan Temme, Madeleine
  Tod, Ellinor Wanzambi, Stephen Wood, and James Wootton.
\newblock Learn quantum computation using qiskit, 2020.

\bibitem{BaldwinQu2019}
C~H Baldwin, K~Mayer, N~C Brown, C~Ryan-Anderson, and D~Hayes.
\newblock Re-examining the quantum volume test: Ideal distributions, compiler
  optimizations, confidence intervals, and scalable resource estimations.
\newblock {\em Quantum}, 6:707, 2022.

\end{thebibliography}

\end{document}